\documentclass[twocolumn,times,tighten,10pt]{aastex631} 
\usepackage[T1]{fontenc}
\usepackage{color}
\usepackage{enumitem}
\usepackage{hyperref}
\usepackage{verbatim}
\usepackage{breqn}
\usepackage{amsmath,amstext,mathrsfs}
\usepackage[all]{hypcap} 
\usepackage{afterpage}  
\usepackage{marginnote}
\usepackage{makecell}

\usepackage{ifxetex,ifluatex}
\usepackage{fixltx2e} 
\usepackage{multirow}
\usepackage{natbib}
\usepackage{graphicx}

\DeclareFontFamily{OMS}{oasy}{\skewchar\font48 }
\DeclareFontShape{OMS}{oasy}{m}{n}{%
     <-5.5> oasy5   <5.5-6.5> oasy6
   <6.5-7.5> oasy7   <7.5-8.5> oasy8
   <8.5-9.5> oasy9   <9.5-> oasy10
   }{}
\DeclareFontShape{OMS}{oasy}{b}{n}{%
    <-6> oabsy5
   <6-8> oabsy7
   <8-> oabsy10
   }{}
\DeclareSymbolFont{oasy}{OMS}{oasy}{m}{n}
\SetSymbolFont{oasy}{bold}{OMS}{oasy}{b}{n}

\DeclareMathSymbol{\smallleftarrow}   {\mathrel}{oasy}{"20}
\DeclareMathSymbol{\smallrightarrow}  {\mathrel}{oasy}{"21}
\DeclareMathSymbol{\smallleftrightarrow}{\mathrel}{oasy}{"24}

\graphicspath{{./}{}}

\shorttitle{MHD mode identification by turbulence statistics}
\shortauthors{Pavaskar et al.}
\received{\today}
\submitjournal{ApJ}

\begin{document}
\title{Diagnostics of magnetohydrodynamic modes in the ISM through synchrotron polarization statistics}

\author[0000-0003-3400-191X]{Parth Pavaskar}
\affiliation{Institute fur Physik und Astronomie Universitat Potsdam, Golm Haus 28, 14476 Potsdam, Germany}
\affiliation{Deutsches Elektronen-Synchrotron DESY, Platanenallee 6, 15738 Zeuthen, Germany}

\author[0000-0003-1683-9153]{Ka Ho Yuen}
\affiliation{Theoretical Division, Los Alamos National Laboratory, Los Alamos, NM 87545, USA}

\author[0000-0003-2560-8066]{Huirong Yan}
\affiliation{Institute fur Physik und Astronomie Universitat Potsdam, Golm Haus 28, 14476 Potsdam, Germany}
\affiliation{Deutsches Elektronen-Synchrotron DESY, Platanenallee 6, 15738 Zeuthen, Germany}
\email{huirong.yan@desy.de}

\author[0000-0003-4147-626X]{Sunil Malik}
\affiliation{Institute fur Physik und Astronomie Universitat Potsdam, Golm Haus 28, 14476 Potsdam, Germany}
\affiliation{Deutsches Elektronen-Synchrotron DESY, Platanenallee 6, 15738 Zeuthen, Germany}


\begin{abstract}

One of the biggest challenges in understanding Magnetohydrodynamic (MHD) turbulence is identifying the plasma mode components from observational data. Previous studies on synchrotron polarization from the interstellar medium (ISM) suggest that the dominant MHD modes can be identified via statistics of Stokes parameters, which would be crucial for studying various ISM processes such as the scattering and acceleration of cosmic rays, star formation, dynamo. In this paper, we present a numerical study of the Synchrotron Polarization Analysis (SPA) method through systematic investigation of the statistical properties of the Stokes parameters. We derive the theoretical basis for our method from the fundamental statistics of MHD turbulence, recognizing that the projection of the MHD modes allows us to identify the modes dominating the energy fraction from synchrotron observations. Based on the discovery, we revise the SPA method using synthetic synchrotron polarization observations obtained from 3D ideal MHD simulations with a wide range of plasma parameters and driving mechanisms, and present a modified recipe for mode identification. We propose a classification criterion based on a new SPA+ fitting procedure, which allows us to distinguish between Alfv\'en mode and compressible/slow mode dominated turbulence. We further propose a new method to identify fast modes by analyzing the asymmetry of the SPA+ signature and establish a new asymmetry parameter to detect the presence of fast mode turbulence. Additionally, we confirm through numerical tests that the identification of the compressible and fast modes is not affected by Faraday rotation in both the emitting plasma and the foreground.


\end{abstract}


\section{Introduction}

The interstellar medium (ISM) is turbulent and magnetized \citep{2010ApJ...725..466C}, spanning over many orders of physical scales, from Au to kpc \citep{Spangler95}. The magneto-hydrodynamic (MHD) turbulence is crucial in governing different physics in the ISM and beyond, from the regulation of heat and thermal phase exchanges in the multi-phase ISM \citep[][]{2003ARA&A..41..241D}, channeling the transport of cosmic rays \citep[CRs,][]{2002PhRvL..89B1102Y, 2004ApJ...614..757Y, YanL08, 2006ApJ...638..811C, 2019PhRvL.123v1103L, 2002cra..book.....S,Kempski2022} and particle acceleration \citep{2004ApJ...611L.101L, 2008ApJ...684.1461Y, lemoine2023nonlinear}, grain dynamics and interstellar chemistry \citep{YLD04, Hirashita:2010ve, Ge2015,Gong2023}, to influencing the formation of cold neutral media \citep{2003ApJ...586.1067H,VDA,cattail, Ho23b} and stars \citep{2007ARA&A..45..565M, Crutcher2012, 2016ApJ...824..134F}. Knowledge of the properties of ISM turbulence is therefore crucial in modeling the ISM and the subsequent star formation processes. 

The complexity of magnetized ISM turbulence along with the limited observational diagnostics, however, restrict us from understanding the physical properties of the turbulence. Typically, the theoretical analysis of MHD turbulence involves the separation of the magnetic field fluctuations into three MHD modes \citep[Alfv\'en, fast and slow, see][]{2002PhRvL..88x5001C, CL03,2020PhRvX..10c1021M}, as each mode exhibits distinct dynamical and statistical properties. One of the biggest challenges in understanding MHD turbulence is the difficulty in identifying the different modes from observational data. For instance, it was proposed by \citet{2002PhRvL..89B1102Y} \citep[see also][] {2004ApJ...614..757Y, YanL08} that the fast mode is much more efficient at accelerating CRs than the Alfv\'en mode. Another example is the presence of the slow mode in multi-phase ISM turbulence which leads to the generation of density features in cold neutral media \citep{Ho23b}. Therefore, characterizing the energy dominance of the different MHD modes is extremely important in understanding some of the unresolved questions regarding the ISM.

Earlier studies \citep{2002PhRvL..88x5001C,  CL03, 2002PhRvL..89B1102Y, 2003ApJ...592L..33Y, 2004ApJ...614..757Y, LP12, KLP17a, VDA, ch5, leakage} indicated that the tensor components of each MHD mode, which are significantly different from one another, are imprinted in the ISM observables. An important development in this direction was the establishment of the Synchrotron Polarization Analysis \citep[SPA,][]{2020NatAs...4.1001Z} technique, suggesting that the energy dominance between that of the Alfv\'en and magnetosonic (MS) modes can be identified via statistics of polarized synchrotron radiation. 

Recent theoretical developments on turbulence statistics (\citealp{leakage}; see also \citealp{MYY2023,ch5}) shed light on analyzing the quantitative energy fractions of the MHD modes from Stokes parameters. The three MHD modes exhibit an exactly orthogonal orientation in 3D space and are integrated along the line of sight in very distinct ways. The additional understanding of the local frame science \citep{leakage} and the impact of magnetic field inclination angle \citep[][also see \S \ref{sec:discussion_synergy}]{MYY2023} reduces the problem of MHD mode-fraction analysis to a simplified geometrical analysis of how the three MHD modes are integrated along the line-of-sight (LOS) in the observed Stokes parameters. Following these developments, the SPA technique requires certain modifications and a more rigorous test of validity. This work addresses the issue by approaching the problem through an MHD mode analysis, and further testing the model on synthetic observations obtained from magnetized turbulence simulations with a significantly wider range of ISM plasma parameters than that considered in the earlier work on the technique. Additionally, we explore the influence of the mean magnetic field geometry and the effect of Faraday rotation through a non-homogeneous foreground media on the observed synchrotron mode signatures, and by extension, the SPA method. This not only facilitates the implementation of the method using real synchrotron data from the observer's standpoint, but it also provides us with the possibility of further modification of the method based on future studies, e.g., further exploration into the asymmetry of observed signatures, etc. Ultimately, we also propose a way to identify the presence of fast modes in the observations, which was not possible in the previous methods. A method that can consistently identify the MHD modes, especially the fast mode, has the potential to significantly improve our understanding of MHD turbulence and its role in astrophysical systems.

In this paper, we give the theoretical description of how the MHD modes can be retrieved from the Stokes parameters in observations using an MHD mode analysis in \S~\ref{sec:theory}. We discuss our numerical approach, simulations, and methodology in \S~\ref{sec:method}, and demonstrate the characteristics of the mode signatures in our technique in \S~\ref{sec:sxx_behavior}. The results of the signature analysis and the mode classification recipe are given in \S~\ref{sec:results_classification}. We compare our techniques to other available methods in \S~\ref{sec:discussion}, discussing the synergies and prospects of our technique. Finally, we conclude our paper in \S~\ref{sec:conclusion}. The appendix of our paper supplements the main text with additional information on our numerical techniques and other relevant concepts.

\section{Theoretical Considerations}
\label{sec:theory}

\subsection{Qualitative description of the Goldreich-Sridhar turbulence theory, and recent developments}
\label{sec:gs95}

The modern description of magnetized turbulence is given by the Goldreich-Sridhar theory \citep{GS95}, which suggests that the magnetized, balanced Alfv\'enic turbulence in the local frame of reference~\citep{2002PhRvL..88x5001C} has a scale-dependent anisotropy of the form of $k_\parallel \propto k_\perp^{2/3}$, which is also proposed to be true for a limited range of imbalanced turbulence \citet{2010ApJ...722L.110B}. For Alfv\'en/pseudo-Alfv\'en modes, the spectral tensor functional form is given by
\begin{equation}
\begin{aligned}
M_{ij}({\bf k}) &= \frac{L_{inj}^{-1/3}}{6\pi}(\delta_{ij}-\frac{k_ik_j}{k^2}) k_\perp^{-10/3} e^{-\frac{L_{inj}^{1/3}|k_\parallel|}{k_\perp^{2/3}}}
\end{aligned}
\label{eq:CL03_tensor_alfven}
\end{equation}
and the tensor for fast modes is given by 
\begin{equation}
\begin{aligned}
M_{ij}({\bf k}) &= \frac{L_{inj}^{-1/2}}{8\pi} \frac{k_i k_j}{k_\perp^2} k^{-7/2} \cos^2\theta \, ,
\end{aligned}
\label{eq:CL03_tensor_fast}
\end{equation}
where $L_{inj}$ is the injection scale and $\theta$ is the angle between ${\boldsymbol k}$ and the magnetic field $B$  \citep[see][]{2002PhRvL..88x5001C, 2002PhRvL..89B1102Y, 2004ApJ...614..757Y}. The model functions (Eq.\ref{eq:CL03_tensor_alfven}, \ref{eq:CL03_tensor_fast}) allow for the analytical study of the orientation of different modes. In general, the spectrum, anisotropy, and the frame (tensor) components (see Appendix of \citealp{leakage} for a summary) contribute to the spectral functions of MHD turbulence.

\begin{figure}
   \centering
   \includegraphics[width=8cm]{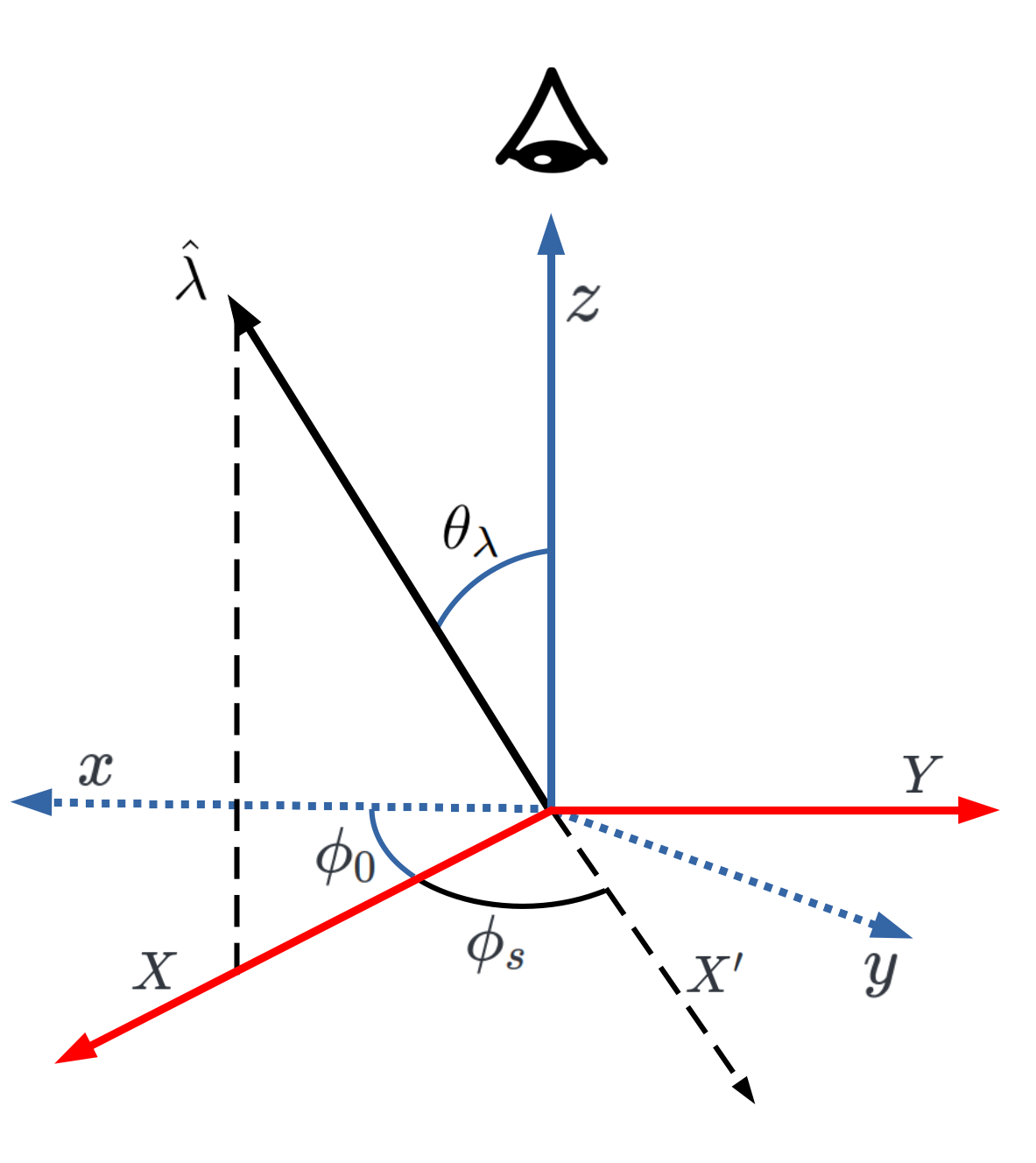}
    \vspace{8pt}
   \caption{A schematic showing the geometry of the system. The $z$-axis represents the LOS. The lowercase $x$ and $y$ show the telescope axes of the observed polarization signatures. The mean magnetic field vector is shown by $\hat{\lambda}$ and $\theta_\lambda$ represents the inclination angle with the $z$ axis).
   The initial Stokes frame ($XY$) is obtained by rotating the telescope frame by an angle equal to the mean polarization angle $\phi_0$. To calculate $s_{xx}(\phi_s)$, the Stokes frame is rotated by angle $\phi_s$ in a step-wise method and the $\mathrm{var}(\epsilon)$ is calculated at each rotated frame $X'Y'$. }
   \label{fig:geometry}
\end{figure}

\subsection{Theoretical basis for the SPA+ signature function ($s_{xx}$)}
\label{sec:toy_model}

We show how the MHD mode fractions can, in principle, be recovered through statistics of observed Stokes parameters in this section. Let us assume that we have a magnetic field with the following configuration, which is comprised of a uniform global field and a turbulent field
\begin{equation}
\label{eq:btotal}
{\bf B} = \bar{\bf B} + {\boldsymbol \delta} \bf{B_A} + {\boldsymbol \delta} \bf{B_C}\, .
\end{equation}
The turbulent field has two components, Alfv\'en mode and compressible (magnetosonic, consisting of fast and slow modes) mode, shown by the subscripts "A" and "C" respectively. This separation is done to see how the mode energy fractions affect the statistics of the observed Stokes
parameters. If ${\bf r}=(x,y,z)$ is the 3D position vector, $z$ points to the observer (LOS), and $\theta_\lambda$ is the mean-field inclination angle, the right-hand side terms in Eq.~\ref{eq:btotal} can be described in the form of:
\begin{equation}
\begin{aligned}
\bar{\bf B} &= \bar{B} \sin \theta_\lambda \hat{z} + \bar{B} \cos \theta_\lambda \hat{x}\\
{\boldsymbol \delta}\bf{B_A}({\bf r}) &= \int d^3 k e^{i{\bf k}\cdot {\bf r}} (C_1) (\hat{k}\times \hat{\bf B})\\
{\boldsymbol \delta}\bf{B_C}({\bf r}) &= \int d^3 k e^{i{\bf k}\cdot {\bf r}} (C_2) (\hat{k}\times (\hat{k}\times \hat{\bf B})) \, ,
\end{aligned}
\label{eq:model}
\end{equation}

where the uniform field lies in the $x-z$ plane. The factors $C_1$ and $C_2$ are used here as generic terms for various two-point statistics (i.e., $|{\bf r}|>0$) that are usually used in other synchrotron analysis methods. In statistical techniques that rely on anisotropy analyses \citep[e.g., ][]{2005ApJ...631..320E,2010ApJ...710..125E,2020NatAs...4.1001Z, MYY2023}, these typically refer to two-point correlation functions or structure functions of various observables. However, since the SPA+ method primarily deals with one-point statistics, the factors serve as straightforward weighting parameters determined by the energy fractions of the Alfv\'en ($C_1$) and compressible ($C_2$) modes. Note that this does not apply to the case of $\theta_\lambda=\pi/2$ since Alfv\'en modes will be projected to zero \citep[see \S~\ref{sec:alfven_sxx}, see also][]{ch5}. For $\theta_\lambda\neq\pi/2$, we can directly derive the Stokes $I,Q,U$ \citep{Heitsch2001}\footnote{Relativistic electron distribution is statistically uncorrelated to ISM turbulence parameters \citep[see ][]{LY18b}.} by

\begin{equation}
\begin{aligned}
I({\bf R}=(x,y)) &= \int dz \left( ({\bf B} \cdot \hat{x})^2 + ({\bf B} \cdot \hat{y})^2 \right)\\
Q({\bf R}=(x,y)) &= \int dz \left( ({\bf B} \cdot \hat{x})^2 - ({\bf B} \cdot \hat{y})^2 \right)\\
U({\bf R}=(x,y)) &= \int dz \left( ({\bf B} \cdot \hat{x})({\bf B} \cdot \hat{y}) \right) \, ,
\end{aligned}
\label{eq:stokes}
\end{equation}
where $z$ denotes the LOS direction. Notice that
\begin{equation}
\begin{aligned}
(\hat{k}\times \hat{\bf B})\cdot \hat{x} & = -\cos\theta_\lambda k_y\\
(\hat{k}\times \hat{\bf B})\cdot \hat{y} & =  -\sin\theta_\lambda k_z+\cos\theta_\lambda k_x\\
(\hat{k}\times (\hat{k}\times \hat{\bf B}))\cdot \hat{x} & = \mu k_x - \cos\theta_\lambda \\
(\hat{k}\times (\hat{k}\times \hat{\bf B}))\cdot \hat{y} & = \mu k_y  \, ,
\end{aligned}
\end{equation}
where $\mu = \hat{k}\cdot \hat{\bf B}$ and $\hat{k} = (k_x,k_y,k_z)$. In principle, the next step involves expressing $\cos\theta_\lambda$ and $\sin\theta_\lambda$ via Rodrigues' rotation and integrating over ${\bf k} = |k|,\mu$, assuming axisymmetric turbulence. However, for our model of the B-field, we can assume that the entire turbulence system is characterized by only one ${\bf k}$ vector (the so-called "one-wave assumption"). In this case, the projection operator here implies $k_z=0$. We can simplify further by replacing the integral $\int dz$ with ${\cal L}_z$, and $\mu = k_x \sin\theta_\lambda + k_z \cos\theta_\lambda \rightarrow k_x \sin\theta_\lambda$. Here, we denote the factor $C_1$ in Eq.~\eqref{eq:model} for Alfv\'en mode as $C_A$, and that for the compressible mode ($C_2$) as $C_C$. Denoting ${\bf K} = (k_x,k_y,0)$, ${\bf R}=(x,y,0)$ and $\phi = \cos({\bf K}\cdot {\bf R})$, we have:
\begin{equation}
\begin{aligned}
I({\bf R}=(x,y)) \approx {\cal L}_z (\bar{B} \sin \theta_\lambda-C_A\cos\theta_\lambda k_y\phi\\+C_C (k_x^2\sin\theta_\lambda - \cos\theta_\lambda)\phi)^2 \\+ (C_A \cos\theta_\lambda k_x\phi + C_C\sin\theta_\lambda k_x k_y\phi)^2\\[2ex]
Q({\bf R}=(x,y)) \approx {\cal L}_z (\bar{B} \sin \theta_\lambda-C_A\cos\theta_\lambda k_y\phi\\+C_C (k_x^2\sin\theta_\lambda - \cos\theta_\lambda)\phi)^2 \\- (C_A \cos\theta_\lambda k_z \phi+ C_C\sin\theta_\lambda k_x k_y\phi)^2\\[2ex]
U({\bf R}=(x,y)) \approx 2{\cal L}_z (\bar{B} \sin \theta_\lambda-C_A\phi\cos\theta_\lambda k_y\\+C_C(k_x^2\sin\theta_\lambda - \cos\theta_\lambda)\phi)\\(C_A \cos\theta_\lambda k_x\phi + C_C \sin\theta_\lambda k_x k_y\phi)
\end{aligned}
\end{equation}

Consequently, one can see why the Stokes parameters are measures of the MHD mode energy fractions as well as the B-field inclination angle $\theta_\lambda$. 
Here we assume that the exchange of energy between the MHD modes is small. While this assumption was justified in previous works based on the numerical simulations of \citet{2002PhRvL..88x5001C}, we find that the mode energies do evolve with time from our simulations, especially with compressively driven turbulence (see Fig.~\ref{fig:mode_frac} in Appendix~\ref{sec:app_efraction}). This means that the stage in the evolution of the modes in the turbulence should also be taken into consideration. This phenomenon will be explored in a separate study. However, our numerical tests show that the identification of the MHD modes based on the modes analysis is possible regardless of these simplifications.

Following \citet{2020NatAs...4.1001Z}, the function we are concerned with is $s_{xx}(\phi_s)$, which is defined as the variance (over ${\bf R}$) of the emissivity of the synchrotron radiation at each rotation of the Stokes axis $\phi_s$. As shown in \citet{2020NatAs...4.1001Z}, the emissivity is given by
\begin{equation}
\epsilon({\bf R};\phi_s) = I + Q_{rot} =I + (Q \cos2\phi_s + U \sin2\phi_s) \, ,
\label{eq:eps}
\end{equation}
where $Q_{rot}$ is the Q parameter in the rotated Stokes frame. Since $\epsilon(\phi)$ has second order $\cos$ and $\sin$ terms, var$(\epsilon)$ is of at most fourth order. This allows us to write a general expression for $s_{xx}=\textrm{var}(\epsilon)$ for arbitrary $\phi_s$ as a Fourier series
\begin{equation}
s_{xx}(\phi_s) = \sum\limits_{n=0}^4 A_n \cos(n\phi_s)+B_n \sin(n\phi_s)
\label{eq:10}
\end{equation}

In the case of $\phi_s=0$ and from Eq.~\ref{eq:eps} (assuming ${\cal L}_x=1$ for simplicity), we have\footnote{$(a+b)^4 \sim a^4 + 6a^2b^2 + b^4$ if odd terms are dropped}
\begin{equation}
\begin{aligned}
\langle \epsilon\rangle^2 &= (\bar{B}^2\sin^2\theta_\lambda \\
&+ \pi(-C_A\cos\theta_\lambda k_y\phi+C_C (k_z^2\sin\theta_\lambda - \cos\theta_\lambda))^2)^2\\
\langle \epsilon^2\rangle &= \bar{B}^4\sin^4\theta_\lambda \\
&+ 6\pi\bar{B}^2\sin^2\theta_\lambda (-C_A\cos\theta_\lambda k_y+C_C (k_z^2\sin\theta_\lambda - \cos\theta_\lambda))^2 \\
&+ 3\pi/4 (-C_A\cos\theta_\lambda k_y+C_C (k_z^2\sin\theta_\lambda - \cos\theta_\lambda))^4 \, .
\end{aligned}
\end{equation}
Subtracting them gives
\begin{equation}
\begin{aligned}
s_{xx}(0) &= (4\pi  \bar{B}^2\sin^2 \theta_\lambda \\
&+ 3\pi/4(-C_A\cos\theta_\lambda k_y+C_C (k_z^2\sin\theta_\lambda - \cos\theta_\lambda))^2) \\ 
&\times(-C_A\cos\theta_\lambda k_y+C_C (k_z^2\sin\theta_\lambda - \cos\theta_\lambda))^2
\end{aligned}
\label{eq:var_e}
\end{equation}
where the first term is the linear signature shown in \citet{2020NatAs...4.1001Z}. It is then clear that the two modes project orthogonally in the plane-of-sky (POS). Effectively, the variance of $(I+Q_{rot})$ carries the information on the mode spectrum and energy fractions $C_A, C_C$ projected distinctively through these weighted terms, which is ultimately embedded in the signature coefficients $A_n$ and $B_n$ in Eq.~\ref{eq:10}.

\section{Method}
\label{sec:method}
\subsection{Simulations}

\bgroup
\setlength\tabcolsep{20pt}
\begin{table*}[t]
\caption{Table of MHD simulations used in the current work. In our simulations, the energy injection rate $\epsilon=0.78$ is fixed to make the turbulent velocity $v_{turb} = 1$. Other default parameters include $L_{box}=1$, $L_{inj}\ge 1/2$, $\langle\rho\rangle=1.$ \label{tab:sim}}
\small
\centering
\begin{tabular}{ccccccc}
\hline
\hline
& & Sonic & & Alfv\'enic &  &  \\
& Sound & Mach & Alfv\'en & Mach & Plasma &  \\
Model Name & Speed & Number & Velocity & Number & Beta &  Resolution \\
& $c_s$ & $M_s$ & $v_A$ & $M_A$ & $\beta$ & $N_x$  \\ 
\hline\hline 
S1 & 0.39 & 3.60 & 1.25 & 0.80 & 0.20 & 576   \\
S2 & 0.62 & 2.20 & 2.00 & 0.50 & 0.20 & 576   \\
S3 & 1.00 & 1.35 & 3.33 & 0.30 & 0.20 & 576   \\
S4 & 3.10 & 1.40 & 10.0 & 0.10 & 0.20 & 576   \\
S5 & 2.80 & 0.35 & 1.25 & 0.80 & 10.0 & 576   \\
S6 & 4.50 & 0.22 & 2.00 & 0.50 & 10.0 & 576   \\
S7 & 7.40 & 0.13 & 3.33 & 0.30 & 10.0 & 576   \\
S8 & 22.0 & 0.04 & 10.0 & 0.10 & 10.0 & 576   \\
\hline
C1 & 1.42 & 0.70 & 6.66 & 0.15 & 0.09 & 576   \\
C2 & 2.00 & 0.50 & 10.0 & 0.10 & 0.10 & 576   \\
C3 & 2.00 & 0.50 & 5.00 & 0.20 & 0.30 & 576   \\
C4 & 2.00 & 0.50 & 2.85 & 0.35 & 1.00 & 576   \\
C5 & 2.00 & 0.50 & 2.00 & 0.50 & 2.00 & 576   \\
C6 & 2.85 & 0.35 & 1.66 & 0.60 & 6.00 & 576   \\
C7 & 3.33 & 0.30 & 1.66 & 0.60 & 8.00 & 576   \\
C8 & 4.00 & 0.25 & 1.50 & 0.65 & 15.0 & 576   \\
\hline
\hline\hline
\end{tabular}
\end{table*}

\egroup

To test the method numerically, we simulate MHD turbulence using the open-source code Athena++ \citep{2020ApJS..249....4S}. We compute time series of three-dimensional, triply periodic, isothermal MHD simulations with impulsive force driving via direct spectral injection. Athena++ uses 3rd-order WENO (Weighted Essentially Non-Oscillatory) discretization, which mitigates spurious oscillations near sharp gradients or discontinuities in the solutions, such as shocks. The time units are normalized to the sound-crossing time $\tau_s=L_{box}/c_s$, where $L_{box}$ is the width of the simulation box and $c_s$ is the isothermal sound speed. We run our simulations for at least 5 $\tau_s$. The other typical parameters for all of our simulation setups are listed in Table~\ref{tab:sim}. The turbulence is driven in such a way that only the eddies at scales $L_{inj} = L_{box}/2$ are subjected to energy injection, which corresponds to driving wavenumbers $|{\bf k_f}|\le 2$. The driving force contains both incompressible (solenoidal) and compressive components controlled by a free parameter $\zeta$ and the forcing function is given by
\begin{equation}
{\bf f} = {\bf f}_{solen} \zeta + {\bf f}_{comp} (1-\zeta) \, ,
\end{equation}
where $\nabla \cdot {\bf f}_{solen} = 0$. To study the behavior of MHD modes in our technique under different plasma environments, we decompose the scalar (density) and vector (velocity, magnetic field) variables in our simulations in the Potential-Alfve\'n-Compressible frame (PAC, see Appendix \ref{sec:app_pca} for details on mode decomposition) to obtain separate datacubes for Alfv\'en and compressible (MS) turbulence. We further separate the fast and slow magnetic field fluctuations from the MS mode by projecting the field onto the respective unit vectors \citep{2002PhRvL..88x5001C}. Collectively, we utilize the mode decomposed simulations to study the signatures of the individual modes in the SPA+ technique and analyze the total magnetic field to cross-check the validity of the method. 

Using the mode decomposed magnetic field simulations, we can also observe how the modes evolve with time. In our simulations, we notice that energy fractions of the MHD modes tend to change substantially as the turbulence evolves over time when we drive the simulation with a compressible forcing term. This particular phenomenon was not observed in earlier studies, since most of the simulations performed previously were driven fully solenoidally. We summarize and discuss this phenomenon in Appendix \ref{sec:app_efraction}.

\subsection{Analysis} \label{sec:analysis}

For each parameter setup from Table~\ref{tab:sim}, we choose snapshots in the time series to analyze the turbulence statistics. It is essential to ensure that the kinetic and magnetic energy densities are fully saturated at the selected time-step, and only the data cubes with saturated turbulence are used to calculate the synthetic synchrotron polarization observations. In general, the synchrotron emission depends on the distribution of relativistic electrons as
\begin{equation}
N_e({\cal E})d{\cal E}\sim {\cal E}^{\alpha} d{\cal E},
\end{equation}
 with the intensity of the synchrotron emission being
\begin{equation}
I_{sync}({\bf X}) \propto \int dz B_{\perp}^\eta({\bf x}) \, , 
\end{equation}
where ${\bf X} = (x,y)$ is the 2D POS vector and $B_{\perp} = \sqrt{B_x^2 + B_y^2} $ is the magnitude of the magnetic field perpendicular to the line of sight in $z$-direction. Generally, $\eta=0.5(\alpha+1)$ is a fractional power law. It has been shown through studies involving synchrotron analysis that the exact value of $\alpha$ does not significantly influence the statistics of $I_{sync}$ and that the assumption of $\alpha=3$ suffices in such a case \citep{2020NatAs...4.1001Z}. For this reason, we will consider the statistics in the limiting case of $\eta =2$ (i.e. $\alpha = 3$) in this study. The synchrotron complex polarization vector with Faraday rotation is given by \citep{Lee2016}:
\begin{equation}
    P_{synch}({\bf R}) = \int dz \, \epsilon_{synch} \, \rho_{rel}B^2e^{2i\left(\theta({\bf R},z)+C\lambda^2\Phi(R,z)\right)}
\end{equation}
where $\epsilon_{synch}$ is the emissivity of synchrotron radiation and
\begin{equation}
    \Phi(R,z) = \int_\infty^z dz' (4\pi)^{-1/2}\rho_{th}({\bf R},z) B_z({\bf R},z) {\rm rad~m^{-2}}
    \label{eq:chap1.frm}
\end{equation} is the Faraday Rotation depth. The $\rho_{rel}$ and $\rho_{th}$ terms are the relativistic and thermal electron densities respectively. The factor C $\approx 0.81$ ~\citep{Kronberg2008,LYLC, malik2020ApJ}. The POS projected magnetic field direction is then given by:
\begin{equation}
    \theta_B = \frac{\pi}{2} + \frac{1}{2} \tan^{-1}_2(\frac{U}{Q})
    \label{eq:chap1.Bangle}
\end{equation}
where $\tan^{-1}_2$ is the 2-argument arc-tangent function. 

Given the above assumptions, the line-integrated Stokes parameters (I, Q, U) at each line-of-sight $\bf R$ on the picture plane can be computed according to Eq.~\ref{eq:stokes}, which gives us 2D Stokes maps for each simulation setup. To take into account the effect of the magnetic field inclination with our LOS, we generate multiple synthetic maps by rotating the simulation box (see Appendix~\ref{sec:app_rotation} for details on the rotation algorithm). Following this step, we compute the $s_{xx}$ parameter similar to \citet{2020NatAs...4.1001Z}. This is done in three steps. In the first step, we choose the region on the 2D polarization map for the calculation of $s_{xx}$, the so-called "analysis spot". The size of this spot is taken to be roughly equal to or less than the coherence scale of turbulence. This naturally implies that the observations have to be performed at a resolution smaller than the coherence length. While the exact coherence scale can be very challenging to estimate observationally, a crude estimate is adequate in the case of the SPA+ method. Such an estimate can be obtained through the measurement of multi-point statistics e.g., second-order (or higher) structure functions (SF) of the observed velocity or intensity data \citep{Cho2019, 2024ApJ...965...65M}. The saturation scale of the SF can be chosen as the upper limit of the size of the analysis spot. On the other hand, the lower limit is simply given by the available resolution of the observations. The spot size can be made arbitrarily small as long as it is inside the inertial scale of turbulence and contains a sufficiently large number of statistics. For synthetic observations, however, we simply choose the energy injection scale in the MHD simulations as the spot size. The second step involves measuring the mean polarization angle from the selected region, which is done using circular statistics\footnote{While circular and linear averaging shows no significant difference in numerically generated synthetic polarization maps, particularly when $M_{A,2D}>1$, we use circular statistics to replicate the method used for real observations}, and rotating the initial Stokes frame (the telescope axis) such that the new ${\bf R}'$ axis in the rotated Stokes frame aligns with the mean polarization angle. In the last step, the new Stokes frame is rotated step-wise in a full circle ($\phi_s \in (0, 2\pi)$) in 360 steps and computes the $s_{xx}$ at each step as $s_{xx}(\phi_s) = \mathrm{var}(\epsilon)(\phi_s)$, where $\epsilon$ is given by Eq.~\ref{eq:eps}. The re-centering done in step 2 ensures that the minima of the $s_{xx}$ function lies in the vicinity of $\pi/2$, which is our area of interest for fitting.

\begin{figure*}
    \centering
    \includegraphics[width=17.5cm]{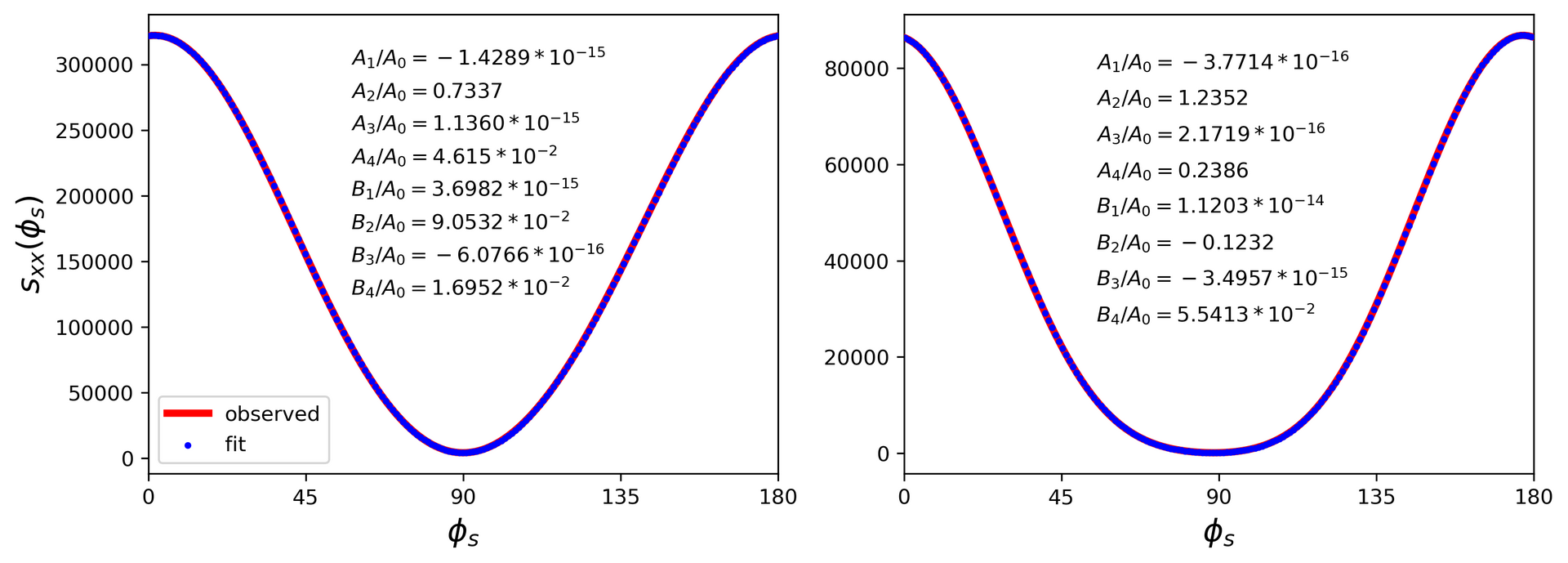}
    \caption{The $s_{xx}$ signatures observed from two simulations (simulation S2 on the left and C5 on the right; see Table~\ref{tab:sim}) are shown in red. Blue dotted lines show the fits to Eq.~\ref{eq:10}. The $sin$ and $cos$ coefficients normalized to $A_0$ are shown for each fit. It is clear that the odd components are negligible and the observed signature can be described by just the even terms.}
    \label{fig:sxx_fit}
\end{figure*}

\subsection{Fitting of the $s_{xx}$ curve}
\label{subsec:fitting}

The resultant $s_{xx}$ function is a sinusoidal-like curve that, from our MHD mode analysis (\S \ref{sec:toy_model}), can be expressed by a fourth-order Fourier series of the rotated frame polar angle $\phi_s$. We can see that this is similar to the fitting function proposed by \citet{2020NatAs...4.1001Z} through their analysis, which is given by 

\begin{dmath}
s_{xx}(\phi_s)_{SPA}  = (a_{xx}\sin^2(\phi_s) + b_{xx} + c_{xx}\sin(\phi_s))\cos^2(\phi_s) \, ,
\label{eq:zhang_sxx_full}
\end{dmath}
where $r_{xx} = a_{xx}/b_{xx}$ was the classification parameter used to identify the modes. Eq.~\ref{eq:zhang_sxx_full} can be rearranged such that

\begin{dmath}
s_{xx}(\phi_s)_{SPA}  = \frac{a_{xx}}{8}(1-\cos(4\phi_s)) + \frac{b_{xx}}{2}(1+\cos(2\phi_s)) +\frac{c_{xx}}{4}(\sin(\phi_s)+\sin(3\phi_s))
\label{eq:zhang_sxx_rearr}
\end{dmath}

where, equating to Eq.~\ref{eq:10}, we get 
\begin{equation}
\begin{aligned}
A_0 &= \frac{a_{xx}}{8} + \frac{b_{xx}}{2}\\
A_2 &= \frac{b_{xx}}{2}\\
A_4 &= -\frac{a_{xx}}{8}\\
B_1 &= B_3 = \frac{c_{xx}}{4} \, 
\label{eq:old_new}
\end{aligned}
\end{equation}
and the rest of the coefficients are equal to zero. However, from preliminary fits of Eq.~\ref{eq:10} to the $s_{xx}$ curves observed from our synthetic polarization maps, we notice that only the even $\sin$ and $\cos$ terms tend to have non-zero coefficients. This is shown in Fig.~\ref{fig:sxx_fit} through examples of $s_{xx}$ observed from two simulations (left and right showing solenoidally and compressively driven respectively) fitted to Eq.~\ref{eq:10}. One can see that the odd terms vanish and the function can be fit using the even terms. Ignoring the odd $\sin$ and $\cos$ terms, Eq.~\ref{eq:10} can be reduced to 
\begin{dmath}
    s_{xx}(\phi_s)  = A_0 + A_2\cos(2\phi_s) + A_4\cos(4\phi_s)+B_2\sin(2\phi_s) + B_4\sin(4\phi_s) \, .
\label{eq:sxx_sim}
\end{dmath}

Essentially, the fit parameters $A_i$ and $B_i$ quantify the features in the shape of the $s_{xx}$ function curve. The coefficients of cosine terms ($A_i$) represent the width of the trough and the slope of the symmetric part of the function near $\phi_s=90^\circ$, whereas the sine coefficients ($B_i$) show the asymmetry, where a negative value represents a left-handed skew and a positive value represents a right-handed skew with respect to the symmetry around $\phi_s=90^\circ$. It can also be noted from Eq.~\ref{eq:old_new} that $A_0$ is not a unique parameter, but rather a combination of $A_2$ and $A_4$, which, along with $B_2$ and $B_4$, are the parameters of interest. More specifically, we can take ratios of the fit parameters to quantify individual features of the $s_{xx}$ curves. We choose our primary classification parameter as $A_4/A_2$ since the width of $s_{xx}$ near $\phi_s=90^\circ$ can be described using the sign of $A_4/A_2$. This parameter, which is essentially identical to the classification parameter $r_{xx}$ used by \citet{2020NatAs...4.1001Z}, can identify the dominating mode from observations based on the unique $s_{xx}$ shapes exhibited by the Alfv\'en and MS modes (See Fig.~\ref{fig:sxx_alf} and Fig.~\ref{fig:sxx_comp}).

In their complete SPA recipe, \citet{2020NatAs...4.1001Z} assumed that the theoretical $s_{xx}$ curves are predominantly symmetric, and chose to ignore the asymmetry term in their fitting function ($c_{xx}$ in Eq.~\ref{eq:zhang_sxx_full}). Consequently, they filtered out all the asymmetric signatures from their analysis of synthetic and observational data. However, from our analysis below, we observe significant asymmetries in the $s_{xx}$ curves, especially in the case of compressively driven turbulence (primarily due to the fast mode e.g. Fig.~\ref{fig:sxx_fast}). In such a case, a symmetry filter would likely filter out a significant portion of the observed data. Furthermore, if the asymmetries are correlated to the modes themselves, a filter would introduce bias in the mode classification scheme. For this reason, we keep the asymmetric $\sin$ terms for our analysis and use Eq.~\ref{eq:sxx_sim} as the fitting function for the observed signatures. We use the parameters $B_2/A_2$ and $B_4/A_2$ to quantify the asymmetry or skewness of the signature. Accordingly, there is no requirement on the degree of asymmetry for observations in our procedure. The reason why the observed $s_{xx}$ signature diverges from symmetry, even when the Stokes frame is re-centered to the POS projected mean magnetic field (which is the mean polarization angle), is not trivial, and it has not been discussed in previous works. While further investigation is required to include the asymmetry in the analytical model, we can use the asymmetry parameter itself as an empirical diagnostic to aid us in the classification of the plasma modes within the scope of this work (see~\S~\ref{sec:results_classification})

\section{Behavior of $s_{xx}$}
\label{sec:sxx_behavior}
To understand exactly how the method can classify the MHD modes in observations, we first take a look at how the $s_{xx}$ function varies when the turbulence is dominated by different MHD modes under different plasma parameters since the $s_{xx}$ can be obtained directly from the Stokes maps without any other inputs. We do this by computing the $s_{xx}$ from the polarization observed from our MHD simulations from Table~\ref{tab:sim} (see section \ref{sec:analysis} on how the synthetic polarization maps are generated). For the coherence scale, we choose a spot equal to $L_{coherence}=L_{inj}$ at the center of the synthetic Stokes map. The application of the method on real observational data requires generating a grid of spots covering the entire regions and repeating steps 2 and 3 from section~\ref{subsec:fitting} for each spot. However, since the simulations are free from any large-scale structures, analysis of an arbitrarily chosen spot is generally good enough for synthetic observations. We decompose and obtain separate datacubes for Alfv\'en and compressible (MS) turbulence, and then compute the $s_{xx}$ signatures from the full simulation as well as the decomposed ones. Finally, we repeat the steps for different plasma parameters (Alfv\'en Mach number for Alfv\'en mode and plasma-$\beta$ for the MS modes) and multiple magnetic field inclination angles ($\theta_\lambda$). We will further discuss the shapes of the $s_{xx}$ curves for Alfv\'en and MS modes separately.

\subsection{Alfv\'en mode}
\label{sec:alfven_sxx}

\begin{figure*}
    \centering
    \includegraphics[width=17.5cm]{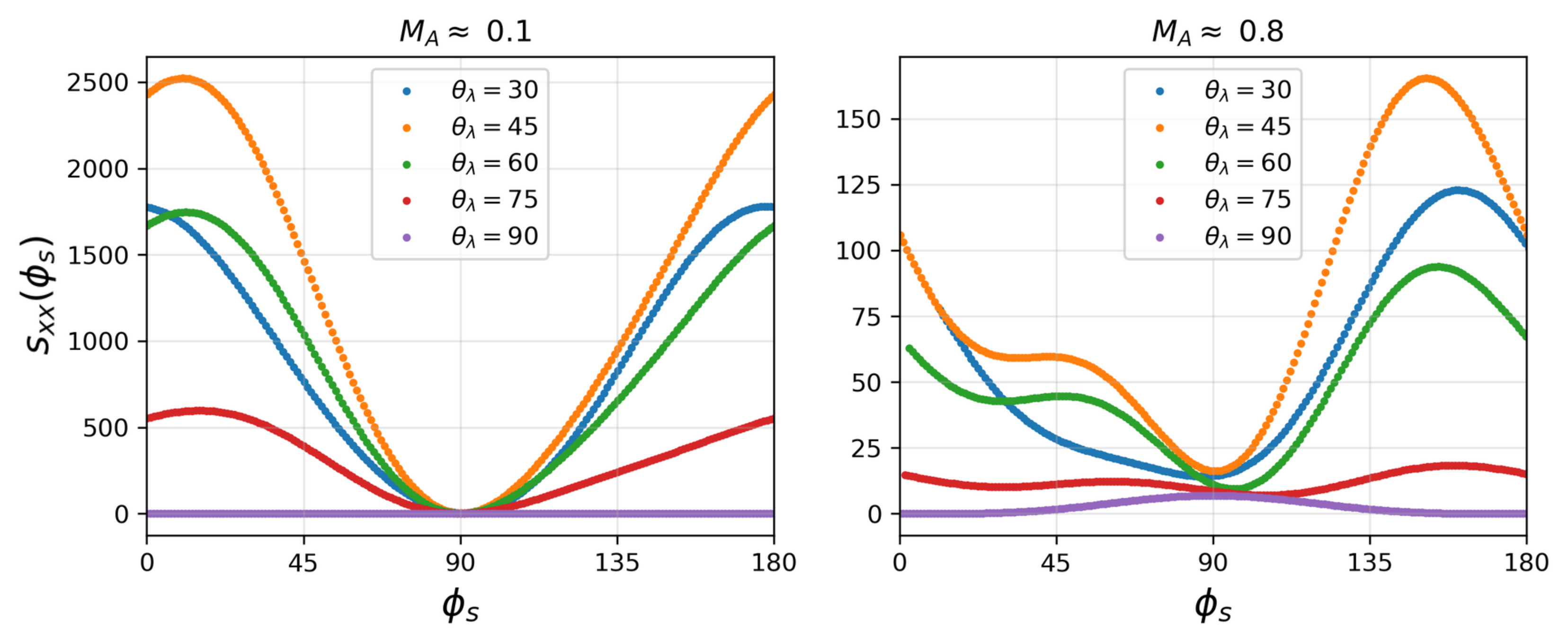}
    \caption{Typical $s_{xx}$ signatures for sub- (left) and trans-Alfv\'enic (right) Alfv\'en mode turbulence decomposed from MHD simulations. The colors represent different magnetic field inclination angles $\theta_\lambda$ relative to the observer.}
    \label{fig:sxx_alf}
\end{figure*}

Before we can study the behavior of the $s_{xx}$ function observed from individually decomposed modes, we first need to ensure that the energy fraction of that particular mode is sufficiently high in the turbulence before the mode decomposition. In the case of Alfv\'en turbulence, we decompose the Alfv\'en mode from fully solenoidally driven turbulence simulations, since solenoidal (divergence-free) driving naturally leads to an incompressible Alfv\'en mode dominant regime. Fig.~\ref{fig:sxx_alf} shows $s_{xx}$ observed from the simulations S4 (left, see Table~\ref{tab:sim}) and S1 (right) after mode decomposition. It should be noted that the amplitude of $s_{xx}$ on the vertical axis is a function of the strength of the mean magnetic field in the simulation as well as the LOS scale, and does not have any significance in the SPA+ technique since the fit parameters are re-normalized after the fit. For the purpose of mode identification, we are primarily interested in the shape of the curve. While the curves in Fig.~\ref{fig:sxx_alf} generally look similar to what was expected in \citet{2020NatAs...4.1001Z}, we would like to point out some important exceptions. A crucial feature is the asymmetry of the curve around $\phi_s=90^\circ$ which does not seem to have any particular dependence on $\theta_\lambda$. It can also be seen that as $\theta_\lambda$ approaches $90^\circ$, the curve starts to get flatter. This can be explained by understanding how the magnetic field fluctuations in Alfv\'en waves project on the POS. It is theoretically expected that pure Alfv\'en waves at $\theta_\lambda = 90^\circ$ are subjected to strong random walk suppression \citep[see, for e.g., Fig.~5 in][]{ch5}. The suppression results in an exponential decrease of polarization angle dispersion in the uncorrelated random walk fashion, i.e. $\delta \phi_{pol,Alf} \sim (L/L_{inj})^{1/2}$, and therefore picking up additional factors of $M_A$ during the estimation of B-field strength. However, \citet{ch5} also discussed that this random walk suppression only happens when $\theta_\lambda$ is exactly $90^\circ$, meaning that a small deviation of the Alfv\'en mode projection from perfectly perpendicular will significantly reduce the random walk suppression issue. In addition to $\theta_\lambda$, the asymmetry in the $s_{xx}$ curve is also sensitive to the Alfv\'en Mach number, where the $s_{xx}$  tends to deviate more from symmetry as $M_A$ approaches 1 (trans-Alfv\'enic limit). We see that as $M_A$ approaches unity, the curves also start to exhibit irregular properties like the shift of the minima, asymmetric peaks, and $\theta_\lambda$ invariance.  This suggests that the SPA+ method is primarily applicable to sub-Alfv\'enic turbulence, and becomes less predictable as the turbulence becomes trans-Alfv\'enic. Generally, for sub-Alfv\'enic turbulence, the $s_{xx}$ seems to be narrow at near $\phi_s=90^\circ$, which suggests that the Alfv\'en mode exhibits $(A_4/A_2)_{Alf}<0$. The signatures are also relatively symmetric, for which we would expect $(B_2/A_2)_{Alf} \approx 0$ and $(B_4/A_2)_{Alf} \approx 0$ (see \S~\ref{sec:results_classification}).

\subsection{Magnetosonic mode}

\begin{figure*}
    \centering
    \includegraphics[width=17.5cm]{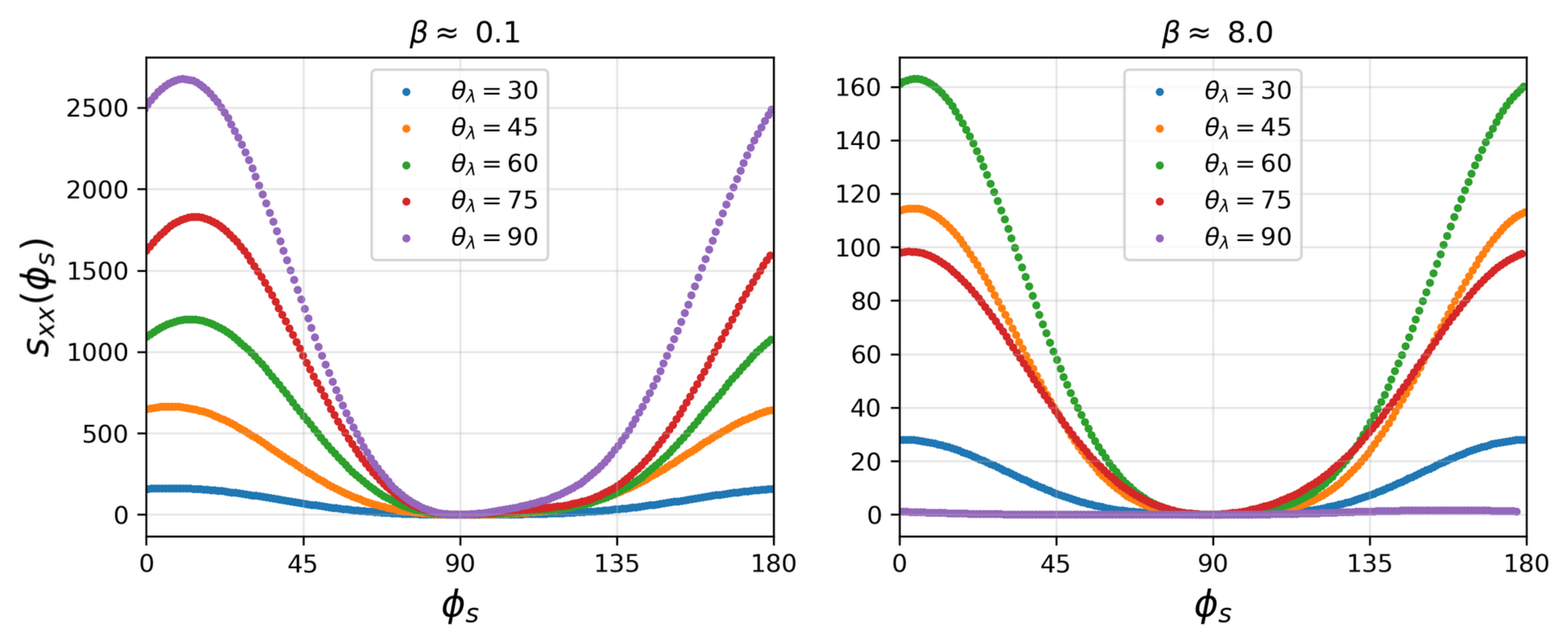}
    \caption{The $s_{xx}$ signatures for magnetosonic mode turbulence decomposed in the P(otential)-C(ompressible)-A(lfv\'en) frame for different $\theta_\lambda$. Left and right panels represent low and high plasma-$\beta$.}
    \label{fig:sxx_comp}
\end{figure*}

Similar to the Alfv\'en case, we compute the $s_{xx}$ from PAC decomposed MS mode turbulence. In this case, however, we use fully compressively driven turbulence simulations to make sure that the energy fraction of MS modes is sufficiently high in the simulation. Unlike the incompressible Alfv\'en modes, the properties of MS modes have a dependence on the plasma-$\beta$. It was reported by \citet{2020NatAs...4.1001Z} that the MS mode classification parameter has a strong $\beta$ dependence. In contrast, we observe little to no dependence of the method parameters on $\beta$. Nevertheless, we present all our results for the compressible modes in two separate plasma-$\beta$ regimes. Fig.~\ref{fig:sxx_comp} shows the $s_{xx}$ observed from the synthetic polarization from the decomposed simulation, showing the cases for low ($\approx 0.1$) and high ($\approx 8$) $\beta$. Given the theoretically predicted behavior of $s_{xx}$ from \citet{2020NatAs...4.1001Z}, the curves look fairly featureless with an expected trough-like shape near $\phi_s=90^\circ$. This particular feature suggests a $(A_4/A_2)_{slow}>0$ for the MS mode. The relatively symmetrical signatures also suggest $(B_2/A_2)_{slow} \approx 0$ and $(B_4/A_2)_{slow} \approx 0$. It can be seen that the MS mode $s_{xx}$ roughly maintains its shape across different plasma-$\beta$ regimes. This is partly expected since, in the case of compressively driven turbulence, the energy fraction of slow modes is much larger than that of fast modes (see Appendix~\ref{sec:app_efraction}), which means that in the total MS mode, the slow mode dominates by a large factor. We can also see this in the form of a very weak dependence of the $s_{xx}$ shape on $\theta_\lambda$. This implies that the MS signature is largely dominated by the slow mode features, and for an analysis of the fast mode, we need to further decompose it from the MS mode.



\subsection{Fast mode}

\begin{figure*}
    \centering
    \includegraphics[width=17.5cm]{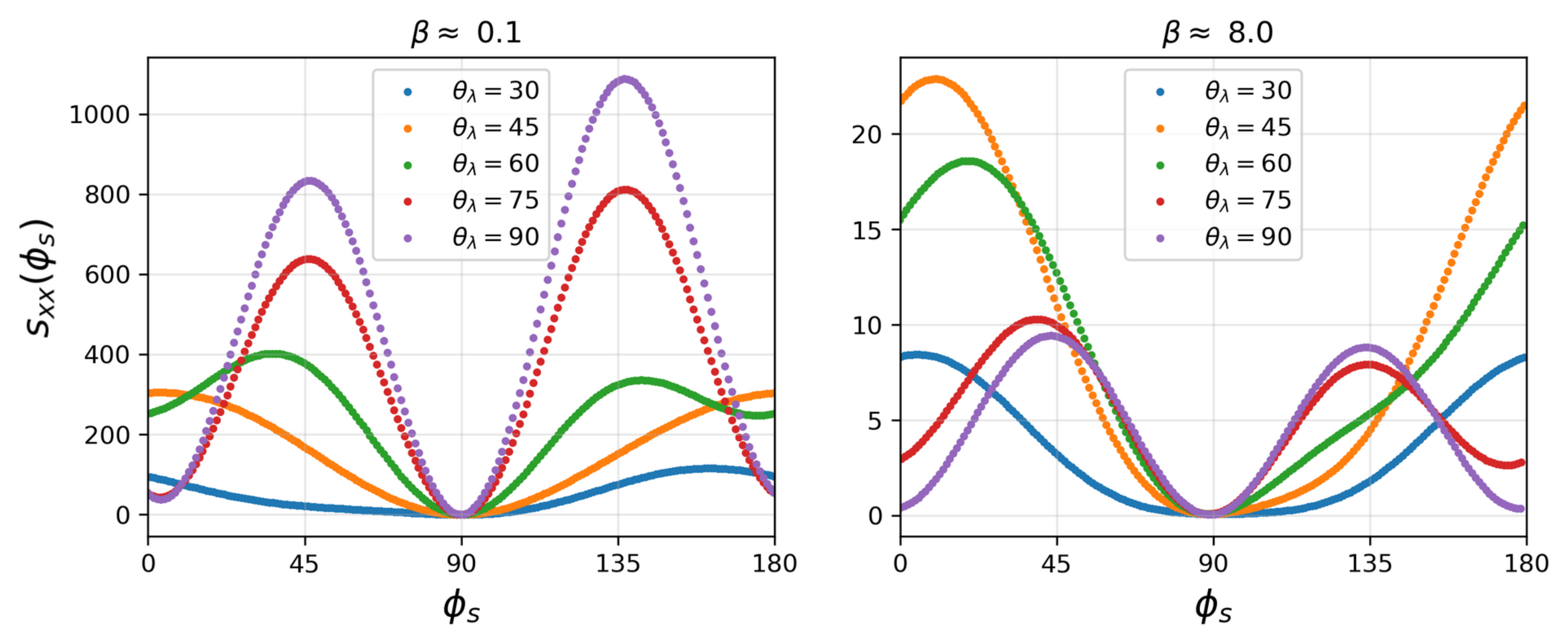}
    \caption{The $s_{xx}$ signatures for fast mode turbulence decomposed in the A(lfv\'en)-S(low)-F(ast) frame for different $\theta_\lambda$. Left and right panels represent low and high plasma-$\beta$.}
    \label{fig:sxx_fast}
\end{figure*}

A limitation of the earlier SPA method is the inability to differentiate between the compressible fast and slow modes, or even make an estimate about the presence of fast modes. This is extremely difficult in observational methods because of the relatively low energy fraction of fast modes in the turbulence. However, the presence of fast modes in interstellar turbulence bears significant implications in gamma-ray astronomy and CR physics \citep{2002PhRvL..89B1102Y,2004ApJ...614..757Y,2008ApJ...677.1401Y,Yan2022rev,Kempski2022}. For this reason, a detection of the presence of fast modes in the ISM could be extremely valuable. To investigate how this could be achieved, we look at the $s_{xx}$ signatures observed from fast mode turbulence. We do this by further decomposing the MS mode into the fast mode in the Alfv\'en-Slow-Fast frame \citep[ASF, ][]{2002PhRvL..88x5001C}. Two of the fast signatures in the high and low $\beta$ regimes are shown in Fig.~\ref{fig:sxx_fast}. From the shapes of the signature functions, it is very clear that fast mode $s_{xx}$ has a very different signature to that of the MS mode while showing no obvious dependence of $\beta$. This further proves that the MS mode signature is largely dominated by the slow mode. The fast mode $s_{xx}$ resembles that of the Alfv\'en mode (see Fig~\ref{fig:sxx_alf}) in the vicinity of $\phi_s=90^\circ$, suggesting $(A_4/A_2)_{fast}<0$. It is also evident that the $s_{xx}$ symmetry changes significantly with B-field inclination. At low $\theta_\lambda$, fast produces a "slow-like" signature (without the peaks away from $\phi_s=90^\circ$) and steadily deviates from it as $\theta_\lambda$ increases. We expect this to reflect in both $(B_2/A_2)_{fast}$ and $(B_4/A_2)_{fast}$ deviating away from 0 as $\theta_\lambda$ increases. It is interesting to note that as $\theta_\lambda$ approaches $90^\circ$, the curve changes into a higher-harmonic sinusoidal-like shape in both $\beta$ regimes, which is also highly asymmetric. This feature is unique to fast modes, and the presence of such a shape asymmetry in an observed signature might suggest the presence of fast mode turbulence with a high B-field inclination angle. Overall, the uniqueness of the shape and asymmetry of the fast mode signature indicates that it should be possible, in principle, to identify the presence of fast modes in turbulence through observations. 

\section{Results}
\label{sec:results}

\subsection{Classification of the dominant mode}
\label{sec:results_classification}

\begin{figure}
    \includegraphics[width=8.5cm]{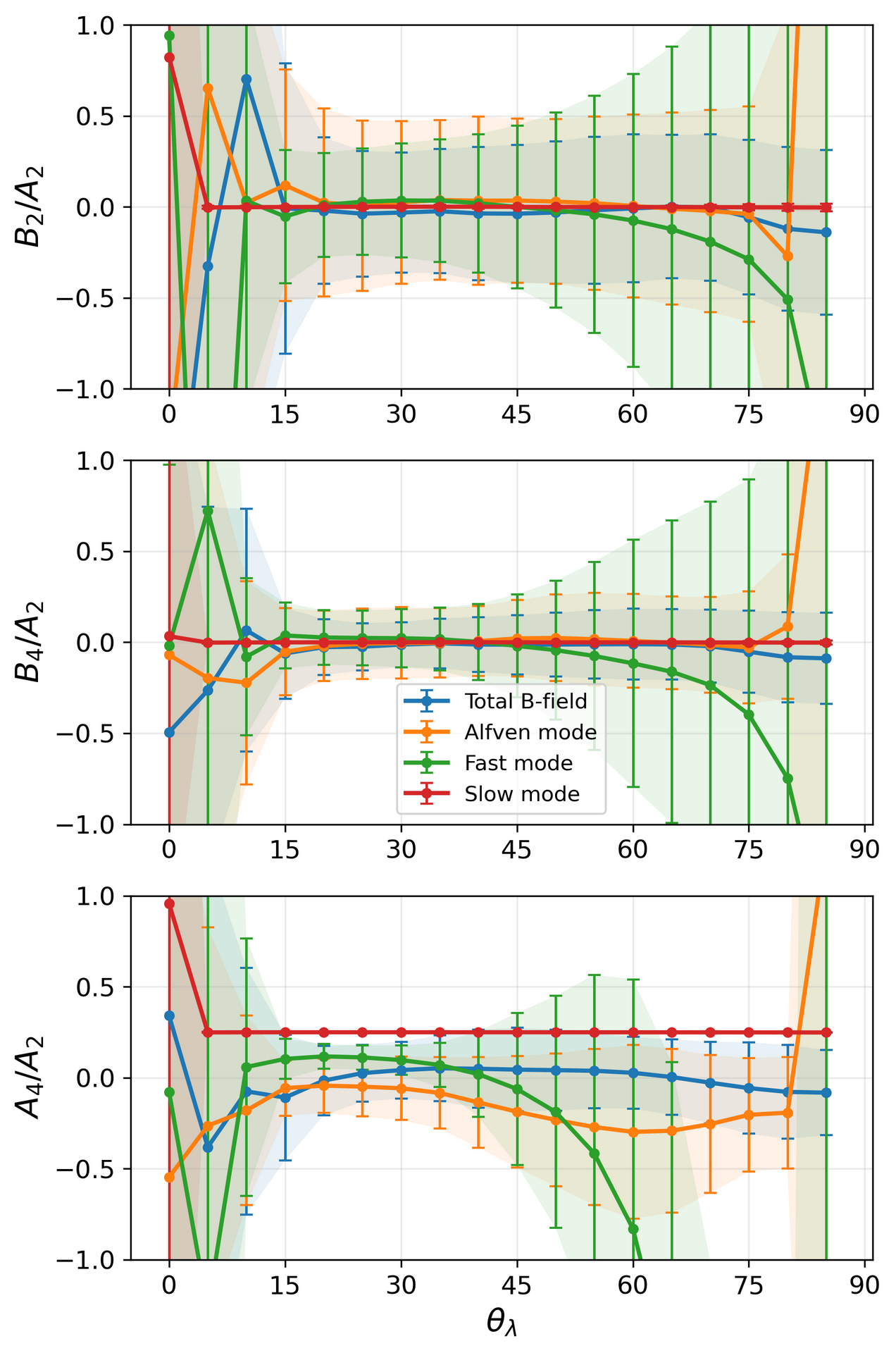}
    \caption{The averaged fit parameters $B_2/A_2$ (top), $B_4/A_2$ (middle), $A_4/A_2$ (bottom) as functions of the B-field inclination angle $\theta_\lambda$, obtained from 24 synthetic synchrotron polarization observations. The synthetic maps are calculated from multiple snapshots in 8 solenoidally driven MHD turbulence simulations. The error bars show $1\sigma$ uncertainties. The parameter $A_4/A_2$ corresponds to the symmetric part of the observed $s_{xx}$ signatures while $B_2/A_2$ and $B_4/A_2$ reflect the asymmetry. The total magnetic field is shown in blue color, and the decomposed Alfv\'en, fast, and slow MHD modes are represented by the colors yellow, green, and red respectively.}
    \label{fig:abb_alfven}
\end{figure}

\begin{figure*}
    \centering
    \includegraphics[width=17.5cm]{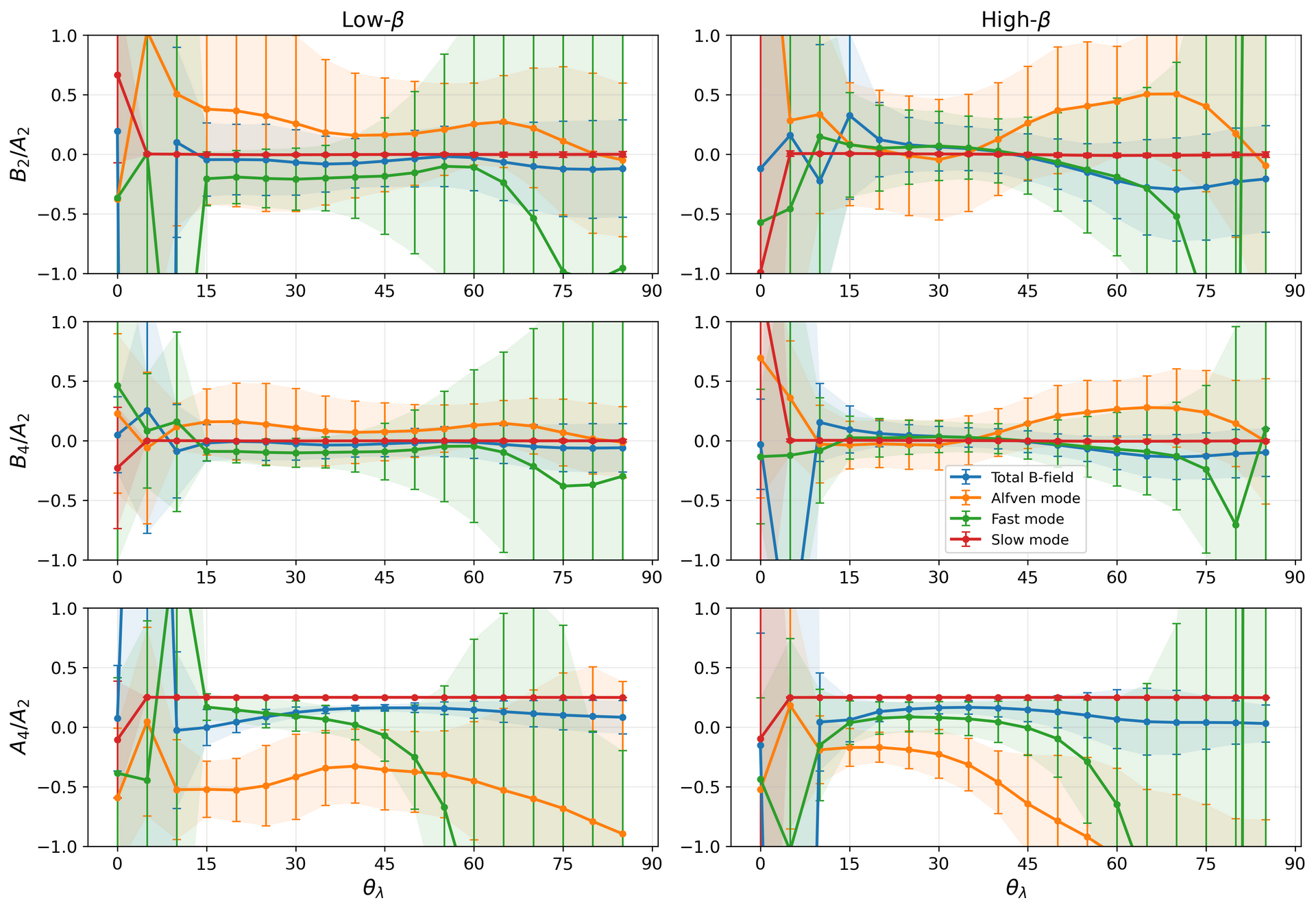}
    \caption{The averaged fit parameters $B_2/A_2$ (top), $B_4/A_2$ (middle), $A_4/A_2$ (bottom) as functions of the B-field inclination angle $\theta_\lambda$, obtained from 24 synthetic synchrotron polarization observations. The maps are obtained from compressively driven MHD simulations. The left and right panels show the parameters obtained from 4 simulations each for low and high plasma-$\beta$ respectively. The total, Alfv\'en, fast and slow magnetic fields are represented by the blue, yellow, green, and red colors respectively.}
    \label{fig:abb_comp}
\end{figure*}

Based on the methodology outlined above, we proceed to describe a recipe to identify the MHD modes from synchrotron polarization observations. Using a large range of MHD turbulence simulations spanning multiple configurations of plasma parameters, we obtain synthetic polarization maps as described in Eq.~\ref{eq:stokes}. While the method in principle is similar to the SPA technique proposed by \citep{2020NatAs...4.1001Z}, we make three notable exceptions. Firstly, we opt to not apply the $s_{xx}$ linearization method to obtain the fit parameters. Instead, we simply perform a fit of Eq.~\ref{eq:sxx_sim} directly to the profile of the observed $s_{xx}$ signature. Secondly, our fitting function incorporates asymmetry terms, which were previously ignored. Lastly, we avoid averaging over the mean B-field inclination angle $\theta_\lambda$ in order to preserve and study the effect of the mean-field geometry and present our fit parameters as functions of $\theta_\lambda$. We perform the SPA+ analysis on all simulations from Table~\ref{tab:sim}, including multiple snapshots of the time evolution of the turbulence. Finally, we separate the results for the Alfv\'en and MS turbulence.

In Fig.~\ref{fig:abb_alfven}, we show the relationship between the fit parameters $B_2$ (top), $B_4$ (middle), and $A_4$ (bottom) and $\theta_\lambda$, averaged across 24 different Alfv\'en mode dominated simulation datacubes (3 time-snapshots for the solenoidally driven simulations S1 - S8 each from Table~\ref{tab:sim}). The three parameters are normalized to $A_2$. The error bars and the shaded area show an uncertainty of $1\sigma$. The blue and orange colors correspond to the total B-field and the decomposed Alfv\'en mode respectively. The fast and slow modes decomposed from the MS mode in the ASF frame are also shown in green and red colors. Upon preliminary inspection, it becomes apparent that when $\theta_\lambda < 15^\circ$, the error in all three observed parameters is too large. This is due to the fact that the mean B-field does not project in the POS when $\theta_\lambda$ is very small. Since the method requires a POS mean field component, we limit our inferences to $\theta_\lambda > 15^\circ$ in the rest of the section.

The parameter $A_4/A_2$ which is shown in the bottom panel is, in principle, similar to the classification parameter $r_{xx}$ in \citet{2020NatAs...4.1001Z}, where the relationship is simply $A_4/A_2=-4r_{xx}$. Their classification scheme was based on the sign of the $r_{xx}$ parameter, where $r_{xx}>0$ (i.e. $A_4/A_2<0$) implies Alfv\'en mode dominance and $-1<r_{xx}<0$ (i.e. $0.25>A_4/A_2>0$) suggests the dominance of MS mode. Since the Alfv\'en mode tends to be negative and the slow mode stays positive with invariance to $\theta_\lambda$ in Fig.~\ref{fig:abb_alfven}, the condition holds for these two modes. However, the fast mode seems to deviate from this rule since $(A_4/A_2)_{fast}>0$ at low $\theta_\lambda$ and it crosses the zero threshold at $\theta_\lambda\approx 35^\circ$. This suggests that when there is a sufficiently high energy fraction of fast modes in the observed turbulence, the symmetry parameter $A_4/A_2$ would not be able to classify the mode signature by itself. However, in our method, we can use the asymmetry parameters $B_2/A_2$ and $B_4/A_2$ to break such a degeneracy. The asymmetry parameters, which are shown in the top and middle panels of Fig.~\ref{fig:abb_alfven}, essentially reflect the degree of asymmetry in the $s_{xx}$ shape. Since both $B_2/A_2$ and $B_4/A_2$ seem to lie close to zero regardless of the $\theta_\lambda$, with the exception of the fast mode, it is clear that when the turbulence is Alfv\'en mode dominated, the signatures are largely symmetrical. The fast mode signature, however, seems to get more asymmetrical as $\theta_\lambda$ increases, which is also reflected in the increasing error bars. This can aid us in the analysis of incompressible turbulence, where a case of $A_4/A_2 < 0$, $B_2/A_2 \approx0$ and $B_4/A_2 \approx0$ implies a strong possibility of Alfv\'en mode dominating the total energy fraction. However, we also see that the $A_4/A_2$ measured from the total magnetic field (shown by the blue color in Fig.~\ref{fig:abb_alfven}) is not less than 0 for all $\theta_\lambda$. This is due to the contribution from the slow mode, which exhibits a significant energy fraction in solenoidally driven simulations (see Fig.~\ref{fig:mode_frac} in Appendix~\ref{sec:app_efraction}).

\begin{figure*}
    \centering
    \includegraphics[width=17.5cm]{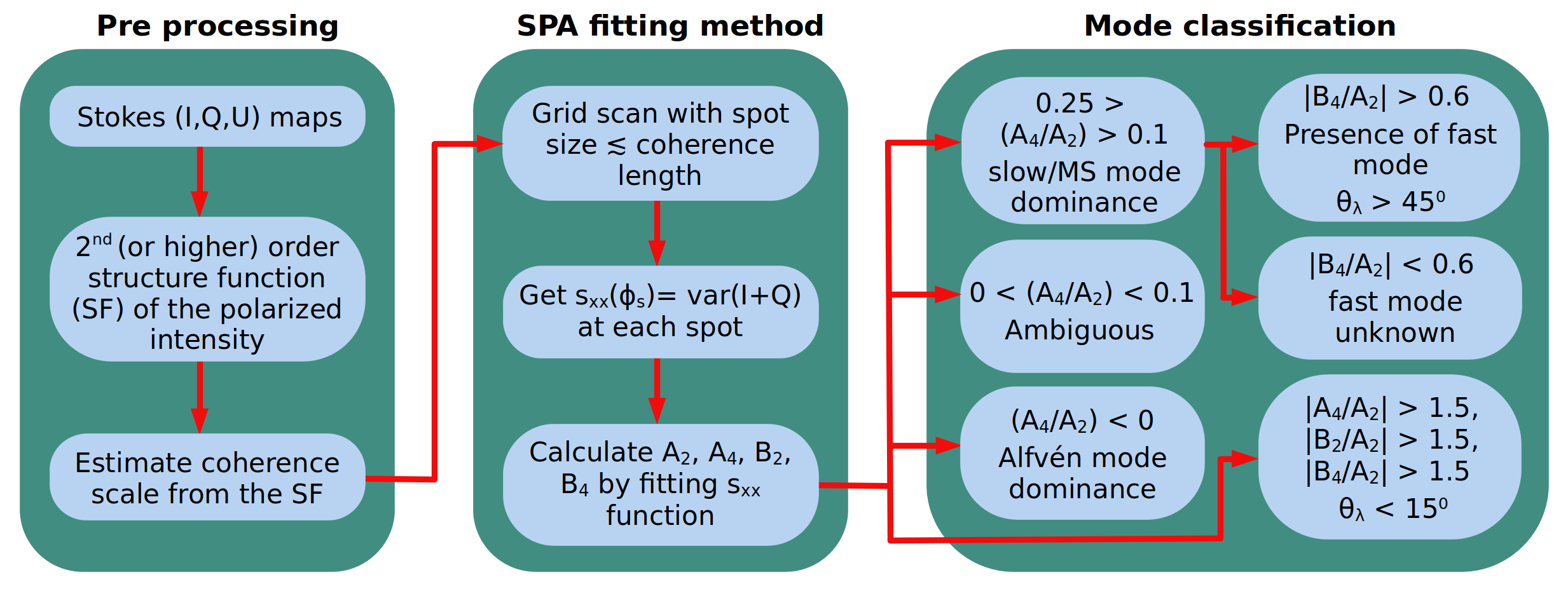}
    \caption{A flowchart showing the full SPA+ mode classification scheme.}
    \label{fig:flowchart}
\end{figure*}

Another consequence of this is that the observations from solenoidally driven simulations are not suitable to make any conclusions for the fast mode signature, since the energy fraction of the fast modes is very low in the total B-field fluctuations in these simulations. For this reason, we use compressively driven simulations to observe how the fast mode signature behaves when the energy of the fast modes in the turbulence is non-negligible. We separate the tests for low and high plasma-$\beta$ for the case of compressible turbulence. Additionally, we notice that simulations driven compressively initially show a large fraction of fast and slow modes but tend to decrease as the turbulence evolves (see Fig.~\ref{fig:mode_frac}). This results in a dominance of Alfv\'en modes in the simulation after approximately 3 sound crossing times. Consequently, only the data cubes where the Compressible modes are dominant are used in the analysis. The fit parameters obtained from compressively driven simulations, averaged across 12 different datacubes for low (3 time-snapshots each from the simulations C1 - C3, left panels) and high (3 time-snapshots each from the simulations C6 - C8, right panels) plasma-$\beta$ each, are shown in Fig.~\ref{fig:abb_comp}. We notice that the asymmetry parameter $A_4/A_2$ behavior (bottom panel) for all three modes is similar to the solenoidal case, where the Alfv\'en and slow modes are also in agreement with \citet{2020NatAs...4.1001Z}. However, similar to the case of fast modes in the solenoidally driven simulations, we cannot make conclusions for the Alfv\'en mode based on compressively driven simulations. For the total B-field, $A_4/A_2$ tends to be positive, which is expected due to the large energy fraction of slow modes in the simulations. We can also see that there is essentially no difference in the symmetry parameter for high and low $\beta$. Generally, we expect a strong slow mode dominance when the turbulence is driven compressively. This suggests that the $A_4/A_2$ is an efficient diagnostic to differentiate between Alfv\'en and slow mode dominance, but is not as effective in detecting fast modes. Even though the fast modes do not gain enough energy to dominate the total magnetic energy, the fast energy fraction is higher in compressible turbulence by a factor of a few than in the case of incompressible turbulence. For the identification of the fast mode, we need to rely on the asymmetry parameters to be able to distinguish it from the Alfv\'en mode. It is interesting to see that, while the average of $B_2/A_2$ and $B_4/A_2$ does not deviate significantly from zero, the widening error bars of the fast mode signature indicates increasing asymmetry of the $s_{xx}$ function as $\theta_\lambda$ increases, especially for the case of low $\beta$. The parameter $B_4/A_2$ shows a clear differentiation between the Alfv\'en and fast modes. In general, while $A_4/A_2<0$ is a valid case for both Alfv\'en and fast modes, the condition $\lvert B_4/A_2 \rvert > 0.6$ would suggest a considerable fast mode energy fraction in the turbulence, along with a large B-field inclination with the LOS. We can use this disparity in the asymmetries of the Alfv\'en and fast mode signatures to identify the presence of fast modes in real synchrotron polarization observations. The asymmetry in addition to the signature of MS mode ($A_4/A_2 > 0.1$) implies a high likelihood of non-negligible energy of fast modes. The overall recipe for the classification between Alfv\'en and slow mode dominance and the identification of fast modes is shown through a flowchart in Fig.~\ref{fig:flowchart}.

\subsection{Estimation of the B-field inclination}
\label{sec:results_inclination}

From Figs. \ref{fig:abb_alfven} and \ref{fig:abb_comp}, we see that the B-field inclination does not always affect the identification of the energy-dominant MHD mode in our technique. Regardless, the fit parameters display some dependence on $\theta_\lambda$. In this section, we discuss the possibility of estimating the B-field inclination in addition to the MHD mode using the modified SPA+ method. As discussed in \S~\ref{sec:results_classification}, we can see that the fast mode can be identified from the asymmetry parameter $B_2/A_2$. However, the asymmetry is only observed at large inclination angles ($\theta_\lambda > 45^\circ$), which means that the identification of the fast mode ($\lvert B_2/A_2 \rvert > 0.6$) along with the MS mode dominance ($A_4/A_2 > 0.1$) suggests a strong possibility of $\theta_\lambda > 45^\circ$. Similarly, from Figs. \ref{fig:abb_alfven} and \ref{fig:abb_comp}, we also notice that the mode identification is not possible when $\theta_\lambda <  15^\circ$ as all three fit parameters show anomalously high or low values. Nevertheless, this can allow us to identify when the B-field is close to alignment with the LOS. Specifically, a large value for the classification parameter ($\lvert A_4/A_2 \rvert > 1.5$) and both the asymmetry parameters ($\lvert B_2/A_2 \rvert > 1.5$, $\lvert B_4/A_2 \rvert > 1.5$) implies a high probability of $\theta_\lambda < 15^\circ$, where the mode identification becomes unreliable.

\section{Discussion}
\label{sec:discussion}

\subsection{The sensitivity of SPA+ on various parameters}

The SPA+ technique is primarily applicable to sub-Alfv\'enic magnetized turbulence. The fit parameters show no sensitivity to the specific value of $M_A$ as long as it is not comparable to unity, or larger. In \S~\ref{sec:sxx_behavior} and \S~\ref{sec:results_classification}, we notice that both the $s_{xx}$ signature and the fit parameters are invariant to the plasma-$\beta$. This means that as long as the observed turbulence is sub-Alfv\'enic with a sufficiently high mean magnetic field inclination ($\theta_\lambda>15^\circ$), our SPA+ technique is robust in its detection of the MHD mode energy fractions. In the case of detection of the fast modes through the asymmetry analysis, the method can also consistently estimate the scenario of a large b-field inclination.

\subsection{Effect of Faraday rotation}
\label{sec:results_faraday}

Since the SPA+ technique relies on measuring the statistical variance of the Stokes parameters in the POS, uniform Faraday rotation (FR) does not affect the classification procedure of the method. To account for non-homogeneous FR, we test the validity of the SPA+ method in two cases, a non-homogeneous foreground rotating screen and the FR in the emitting plasma, and analyze the resulting synthetic polarization maps in the SPA+ framework. Our tests indicate that non-uniform FR, whether in the emitting plasma or foreground, tends to impact the method's fit parameters similarly. Specifically, it leads to an underestimation of the measured $A_4/A_2$, while $B_2/A_2$ and $B_4/A_2$ remain largely unaffected. We illustrate the effect of FR on $A_4/A_2$ from a foreground FR screen in Fig.~\ref{fig:faraday_plot} (Appendix \ref{sec:app_fr}). An underestimation of the $A_4/A_2$ means that in the case of a large FR error, the MS modes will be misidentified as Alfv\'en modes. Consequently, the identification of Alfv\'en modes might be unreliable when the FR effect is large, but this demonstrates the reliability of the MS modes identified by the method, as no Alfv\'en modes will be mislabeled as MS modes. Furthermore, the lack of any significant variation in $B_2/A_2$ and $B_4/A_2$ implies that the FR does not affect the asymmetry of the $s_{xx}$ signature, thereby confirming that the identification of the fast modes in the presence of FR is also robust at least in the case of smooth turbulent foreground. The impact of intermittent structures may be removed with Faraday tomography, which we will address in a future study.

Given that the primary aim of the SPA+ method is to consistently identify the MS modes, and particularly the fast modes, from observational data, we can assert that the technique remains entirely robust even in the presence of non-homogeneous FR.

\subsection{Synergy to previous methods and other synchrotron statistical techniques}
\label{sec:discussion_synergy}

Identification of the MHD modes in the ISM, and especially the presence of the fast mode, is of utmost importance in the study of various processes such as CR transport and acceleration. The previous SPA method for the determination of the MHD modes dominating the energy fraction in the plasma, distinguishing between Alfv\'enic and magnetosonic (compressible) modes, but could not distinguish the fast mode through observations. The present paper seeks to address this limitation by providing a method to determine the presence of fast modes through asymmetry analysis of the mode signature.

The knowledge of the dominant mode fraction from the SPA+ method can be effectively integrated with recently developed techniques in ISM studies. For example, the Velocity Gradient Technique \citep[VGT,][]{YL17a} has made advancements in distinguishing between media dominated by slow/Alfv\'en modes, and those dominated by fast modes. In a medium comprising a mixture of these modes, the absence of mode energy fraction information can introduce a $90^\circ$ ambiguity, similar to \cite{1981ApJ...243L..75G} effect, regarding the actual direction of the magnetic field. This ambiguity is also observed in the synchrotron gradient variant with strong Faraday rotation \citep{LY18b}, where the Stokes parameters no longer provide reliable measures of the magnetic field direction. The degeneracy can be broken only with precise measurement of the polarization of spectral lines through the Ground State Alignment (GSA) effect \citep{YanL06, YanL07, 2008ApJ...677.1401Y, GSA_obs20, Pavaskar2023} so far. Hence, by utilizing our current technique, we can reveal the dominance of a specific mode in the magnetic field, thereby resolving the $90^\circ$ ambiguity and providing a more accurate determination of the magnetic field direction.

Recently, a method for simultaneous retrieval of the line-of-sight angle $\theta_{\lambda}$ and mode fraction was proposed by ~\citep{MYY2023} based on the mapping theory of MHD turbulence statistics~\citep{LP12, leakage}. This approach, known as "Y-parameter analysis", utilizes two-point observable statistics to examine the anisotropies in the magnetic fluctuations. The method relies on the observable quantities $I+Q\propto B_x^2$ and $I-Q\propto B_y^2$, where $B_x$ and $B_y$ represent the plane-of-sky components of the magnetic field. The Y-parameter, defined as the ratio of the anisotropy of $D_{I+Q}$ to the anisotropy of $D_{I-Q}$, captures the characteristics of the embedded magnetic field fluctuations induced by turbulence. It is expressed as:

\begin{equation}
Y = \frac{\text{Anisotropy}(D_{I+Q})}{\text{Anisotropy}(D_{I-Q})} = \frac{R_{v}/R_{h}(D_{I+Q})}{R_{v}/R_{h}(D_{I-Q})},
\label{eq_y}
\end{equation}

where $R_{v}$ and $R_{h}$ represent the extent of correlation function distribution in the vertical and horizontal direction and B$_{POS}$ direction defines the vertical axis. In their study, \citet{MYY2023} applied this statistical technique to simulated MHD cubes to quantify its effectiveness. They established a statistical criterion of ${\rm Y} \sim 1.5\pm0.5$ to identify the dominant fraction of MHD turbulence modes, with ${\rm Y} > 1.5$ indicating the Alfv\'en mode dominance and ${\rm Y} < 1.5$ indicating compressible mode dominance. Interestingly, the Y-parameter exhibited contrasting trends, either increasing or decreasing, with respect to the mean field inclination angle $\theta_\lambda$ for Alfv\'en and compressible turbulence modes. This characteristic enables the utilization of these statistical measures to infer the magnetic field's inclination relative to the line of sight in turbulent environments such as the ISM and the ambient regions of pulsar wind nebulae (PWNe). The method holds a strong synergy with the SPA+ method owing to the following important facts. Firstly, the SPA+ analysis is able to validate the mode dominance estimated by the Y-parameter approach. Secondly, and possibly more importantly, an agreement of a measurement of a high $\theta_\lambda$ through the asymmetry analysis in SPA+ and the Y-parameter recipe could be a robust confirmation of the presence of fast modes in the observations. Furthermore, the identification of compressible modes, including the fast mode, in the SPA+ method is not influenced by FR, making it highly complementary to the Y parameter technique.

\subsection{Implications to cosmic ray studies}
\label{sec:discussion_cr}

The scattering efficiency for the fast mode was predicted by \citet{2002PhRvL..89B1102Y}, showing a significant increase by orders of magnitude compared to that of the Alfv\'en mode. This is due to the fact that the fast modes are highly isotropic \citep{CL03,2020PhRvX..10c1021M}, unlike the Alfv\'en modes which show scale-dependent anisotropy \citep{GS95}. This means that CR acceleration is most effective when a sufficient number of fast modes are present in the magnetized turbulence system. Our current method, based on asymmetry analysis, allows us to detect the presence of a relatively high energy fraction of fast modes in the plasma. This discovery holds crucial implications for understanding both CR scattering and acceleration \citep[see also][]{2004ApJ...614..757Y,2006ApJ...638..811C,2008ApJ...684.1461Y}. Our detection of the fast mode can be cross-checked with studies of CR energy distributions and gamma-ray observations to provide a more comprehensive understanding of how ISM interacts with the CRs \citep[see, for example,][]{Yan2022rev,Kempski2022}. Such observations can potentially also shed light on the physics underlying some of the unexplained high-energy CR sources.

\section{Conclusion}
\label{sec:conclusion}

In this paper, we propose a modified technique (SPA+) for diagnosing the energy-dominant plasma modes through Stokes parameter statistics built upon the existing SPA technique \citep{2020NatAs...4.1001Z}. Particularly, we show that it is possible to detect the presence of fast modes through the analysis of the asymmetry of the SPA+ signature. To summarize our findings:
\begin{enumerate}
\item From an MHD mode analysis, we see that the MHD mode vectors play a dominant role in deciding how the polarization signals are integrated along the line of sight. Particularly, the Alfv\'en and magnetosonic modes are projected completely differently on the plane of the sky owing to their orthogonal 3D orientation, which is reflected in the varying behavior of the $s_{xx}$ signature. 
\item Analysis of the shape of the $s_{xx}$ curve allows us to determine whether the fluctuations are Alfv\'en-like or compressible-like (which suggests a high likelihood of the energy dominance of slow modes).
\item Quantification of the asymmetry of $s_{xx}$ through curve-fitting makes it possible to detect the presence of fast modes, which show a significantly larger asymmetry compared to that of the Alfv\'en and slow modes. The detection of a large signature asymmetry also corresponds to a large mean magnetic field inclination with respect to the line of sight.
\item The mode classification framework provides a robust diagnosis irrespective of plasma-$\beta$ and $M_A$, as long as the turbulence is sub-Alfv\'enic. The method is applicable for all magnetic field geometries as long as the inclination angle $\theta_\lambda > 15^\circ$.
\item The identification of the compressible modes, and particularly the fast mode, is not influenced by Faraday rotation in both the emitting plasma and the foreground.
\item  The SPA+ method can potentially estimate the mean magnetic field inclination $\theta_\lambda$ in two cases: $\theta_\lambda > 45^\circ$ and $\theta_\lambda < 15^\circ$ from the signature asymmetry analysis.

\end{enumerate}

{\noindent {\bf Acknowledgment}
PP, HY, and SM gratefully acknowledge the computing time granted by the Resource Allocation Board and provided on the supercomputer Lise and Emmy at NHR@ZIB and NHR@Göttingen as part of the NHR infrastructure. The numerical calculations for this research were conducted with computing resources under the project bbp00062 (2022), bbp00065 \& bbp00066 (2023). The research by KHY was supported by the Laboratory Directed Research and Development program of Los Alamos National Laboratory under project number(s) 20220700PRD1. This research used resources provided by the Los Alamos National Laboratory Institutional Computing Program, which is supported by the U.S. Department of Energy National Nuclear Security Administration under Contract No. 89233218CNA000001. This research also used resources of the National Energy Research Scientific Computing Center (NERSC), a U.S. Department of Energy Office of Science User Facility located at Lawrence Berkeley National Laboratory, operated under Contract No. DE-AC02-05CH11231 using NERSC award FES-ERCAP-m4239 (PI: KHY, LANL).

} 

\bibliography{refs.bib} 
\bibliographystyle{aasjournal}

\clearpage

\appendix

\section{Time evolution of mode energy fractions in ATHENA++}
\label{sec:app_efraction}
\begin{figure}[h]
    \centering
    \includegraphics[width=17.5cm]{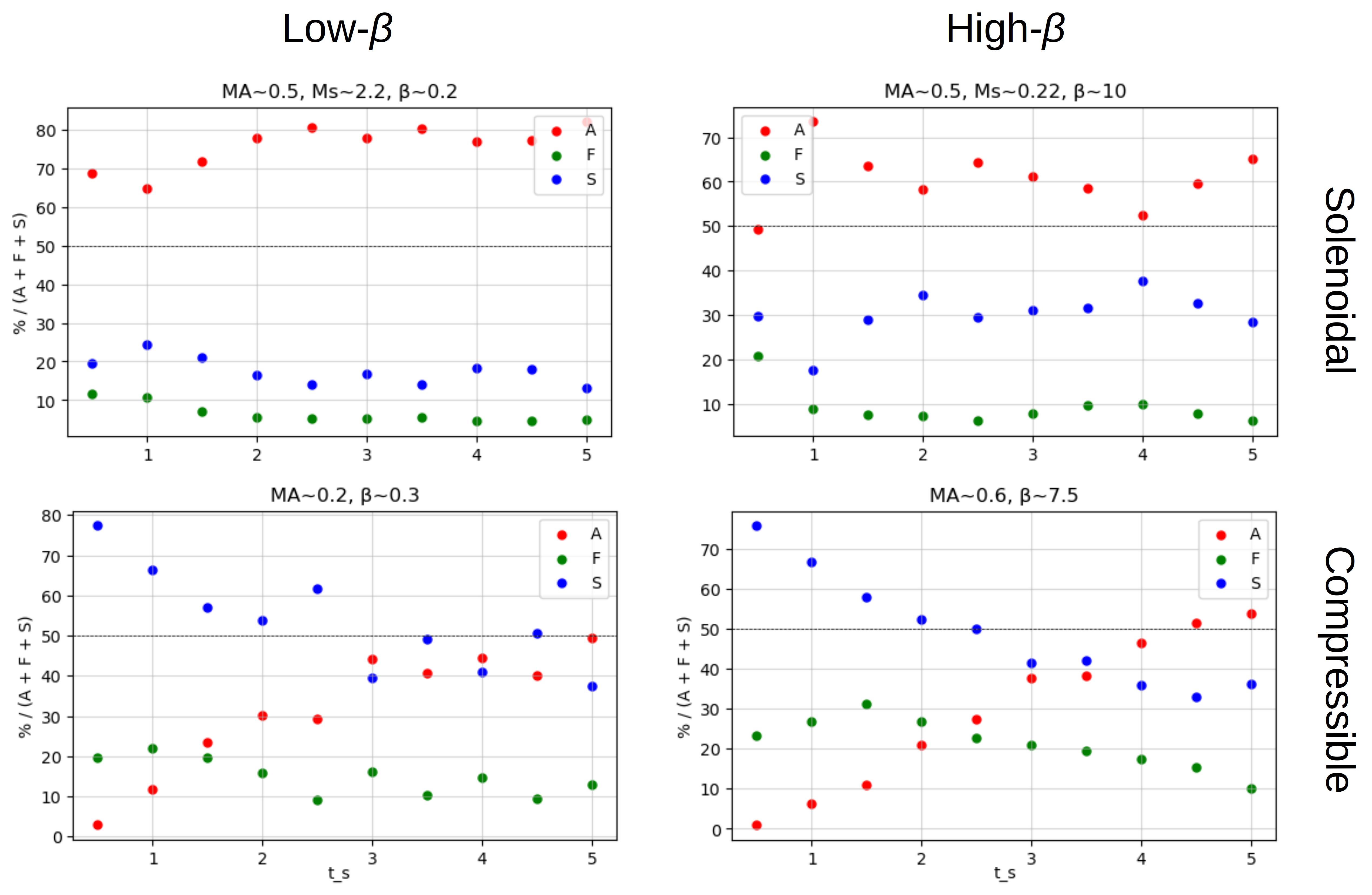}
    \caption{A set of figures showing the energy fraction evolution for the three MHD modes for 4 selected simulations divided into plasma-$\beta$ regimes and driving mechanisms. The x-axis is in units of sound crossing time.}
    \label{fig:mode_frac}
\end{figure}

In this section, we show examples of the energy evolution of each MHD mode, decomposed in the ASF frame \citep{CL03} for a few selected MHD simulations in Fig.\ref{fig:mode_frac}. Four simulations (S2, S6, C3, C7, see Table\ref{tab:sim}) with different driving strength $\zeta \in (0,1)$ and plasma $\beta$ are chosen. As expected, solenoidal driving typically leads to Alfv\'en mode dominated turbulence, while compressible driving leads to the dominance of MS modes (primarily the slow mode)\citep[see also][]{2020PhRvX..10c1021M}.
 
In the case of solenoidally driven simulations, the kinetic and magnetic energies saturate at $\tau_s\approx 2.5$. Following the saturation, the mode energies appear to evolve with approximately constant fractions, which is the expected behavior. Compressively driven turbulence, on the other hand, exhibits a curious time evolution of the mode energies post magnetic energy saturation ($\tau_s \approx 1$). The Alfv\'en mode energy rises rapidly, complemented by the rapid decrease in the slow and fast mode energies, to the point where a situation similar to that of the solenoidally driven turbulence arises, after which the mode energies evolve in a steady state.

\section{Rotation algorithm}
\label{sec:app_rotation}

In our simulations, the initial mean magnetic field direction is always parallel to the z-direction ($\theta_\lambda = 0^\circ$ when the z-axis is the LOS). To produce more samples of mean field orientations, we perform the 3D Rodrigues' rotation algorithm\footnote{\url{https://www.github.com/doraemonho/LazRotationDev}}.  The rotation matrices are defined as :
\begin{equation}
    \begin{aligned}
    {\bf \hat{T}}_x &= \left[\begin{array}{ccc}
        1 & 0 & 0 \\
        0 & \cos(\theta_x) & -\sin(\theta_x) \\
        0 & \sin(\theta_x) &  \cos(\theta_x) \\
    \end{array}\right]\\
    {\bf \hat{T}}_y &= \left[\begin{array}{ccc}
        \cos(\theta_y) & 0 & \sin(\theta_y) \\
        0 & 1 & 0 \\
        -\sin(\theta_y) & 0  &  \cos(\theta_y) \\
    \end{array}\right]\\
    {\bf \hat{T}}_z &= \left[\begin{array}{ccc}
        \cos(\theta_z) & -\sin(\theta_z) & 0 \\
        \sin(\theta_z) &  \cos(\theta_z) & 0 \\
        0 & 0&  1\\
    \end{array}\right]\\
    \end{aligned}
    \label{eq:euler_rotation}
\end{equation}
where we can write the rotation matrix ${\bf \hat{T}} = {\bf \hat{T}}_x{\bf \hat{T}}_y{\bf \hat{T}}_z$, and $\theta_{x,y,z}$ are desired rotation angles along the x,y,z axes respectively.

\section{Mode decomposition}
\label{sec:app_pca}

Decomposition of the simulated magnetic field in the P(otential)-A(lfven)-C(ompressible) components is performed by projecting the magnetic field Fourier component onto the mode bases in the PAC frame given by

\begin{equation}
    \begin{aligned}
        \hat{\zeta}_P(\hat{\bf k},\hat{\lambda}) &= \hat{\bf k} \\
        \hat{\zeta}_A(\hat{\bf k},\hat{\lambda}) &\propto \hat{\bf k} \times \hat{\lambda} \\
        \hat{\zeta}_C(\hat{\bf k},\hat{\lambda}) &\propto \hat{\bf k} \times (\hat{\bf k} \times \hat{\lambda})
    \end{aligned}
    \label{eq:pac}
\end{equation}
where the mean magnetic field unit vector is given by $\hat{\lambda}$. The PAC frame has its special advantage since the sampling of ${\bf k}$ is usually complete in $d\Omega_k$. That means we have the freedom to fix ${\bf k}$ despite the changes in other unit vectors. We can write an arbitrary vector in the Fourier space as:
\begin{equation}
    {\boldsymbol \zeta}({\bf k}) = C_P \hat{\bf k} + C_C \frac{(\hat{\bf k}\times (\hat{\bf k} \times \hat{\lambda}))}{ |\hat{\bf k} \times \hat{\lambda}|} + C_A \frac{(\hat{\bf k} \times \hat{\lambda})}{|\hat{\bf k} \times \hat{\lambda}|}
\end{equation}
which we will name the unit vector $\zeta_{P,A,C}$ for the definition of symbols. The projection in Fourier space of the magnetic field vectors from the simulations along the unit vectors $\zeta_{P,A,C}$ gives us the decomposed magnetic fields with fluctuations arising from the respective modes. 
For the decomposition of fast and slow modes in the (A)lfv\'en--S(low)-F(ast), we use the following bases in the case of adiabatic or isothermal MHD \citep[see Appendix A in][for a detailed discussion]{CL03} :
\begin{equation}
    \begin{aligned}
    \zeta_A(\hat{\bf k},\hat{\lambda}) &\propto \hat{\bf k} \times \hat{\lambda}\\
    \zeta_S(\hat{\bf k},\hat{\lambda}) &\propto (-1 +\alpha-\sqrt{D}) ({\bf k}\cdot \hat{\lambda}) \hat{\lambda}  \\&+ (1+\alpha - \sqrt{D}) ( \hat{\lambda} \times ({\bf k}\times \hat{\lambda})) \\
    \zeta_F(\hat{\bf k},\hat{\lambda}) &\propto (-1 +\alpha+\sqrt{D}) ({\bf k}\cdot \hat{\lambda}) \hat{\lambda}  \\&+ (1+\alpha + \sqrt{D}) ( \hat{\lambda} \times ({\bf k}\times \hat{\lambda})) \\
    \end{aligned}
    \label{eq:cho}
\end{equation}
for Alfv\'en, slow and fast modes respectively, where $\alpha = \beta\Gamma/2$, $D=(1+\alpha)^2- 4\alpha\cos ^2\theta_\lambda$, and $\theta_\lambda$ is the angle between $\hat{\bf k}$ and $\hat{\lambda}$. The plasma-$\beta\equiv P_{gas}/P_{mag}$ measures the plasma compressibility and $\Gamma =\partial P/\partial \rho$ is the polytropic index of the adiabatic equation of state ($\Gamma=1$ for the case of isothermal equation of state). The presence of $\hat{\bf k}$ suggests that the direction of the three mode vectors change as ${\bf k}$ changes. In this scenario, the perturbed quantities, e.g. for the velocity fluctuations ${\bf v}_1 = {\bf v}-\langle {\bf v}\rangle$ can be written as:
\begin{equation}
    {\bf v}_1({\bf r}) = \int_{\bf k}  e^{i{\bf k}\cdot {\bf r}} \sum_{X\in A,S,F} F_{0,X}({\bf k})F_{1,X}({\bf k},\hat{\lambda})  C_X \zeta_X(\hat{\bf k},\hat{\lambda}) \, d^3 {\bf k}
    \label{eq:2}
\end{equation}

The magnetic field can be obtained through a similar projection as in the case of the PAC frame, where the compressive mode will be further decomposed into fast and slow modes.

\section{Numerical tests of Faraday Rotation}
\label{sec:app_fr}

\begin{figure*}
    \centering
    \includegraphics[width=17.5cm]{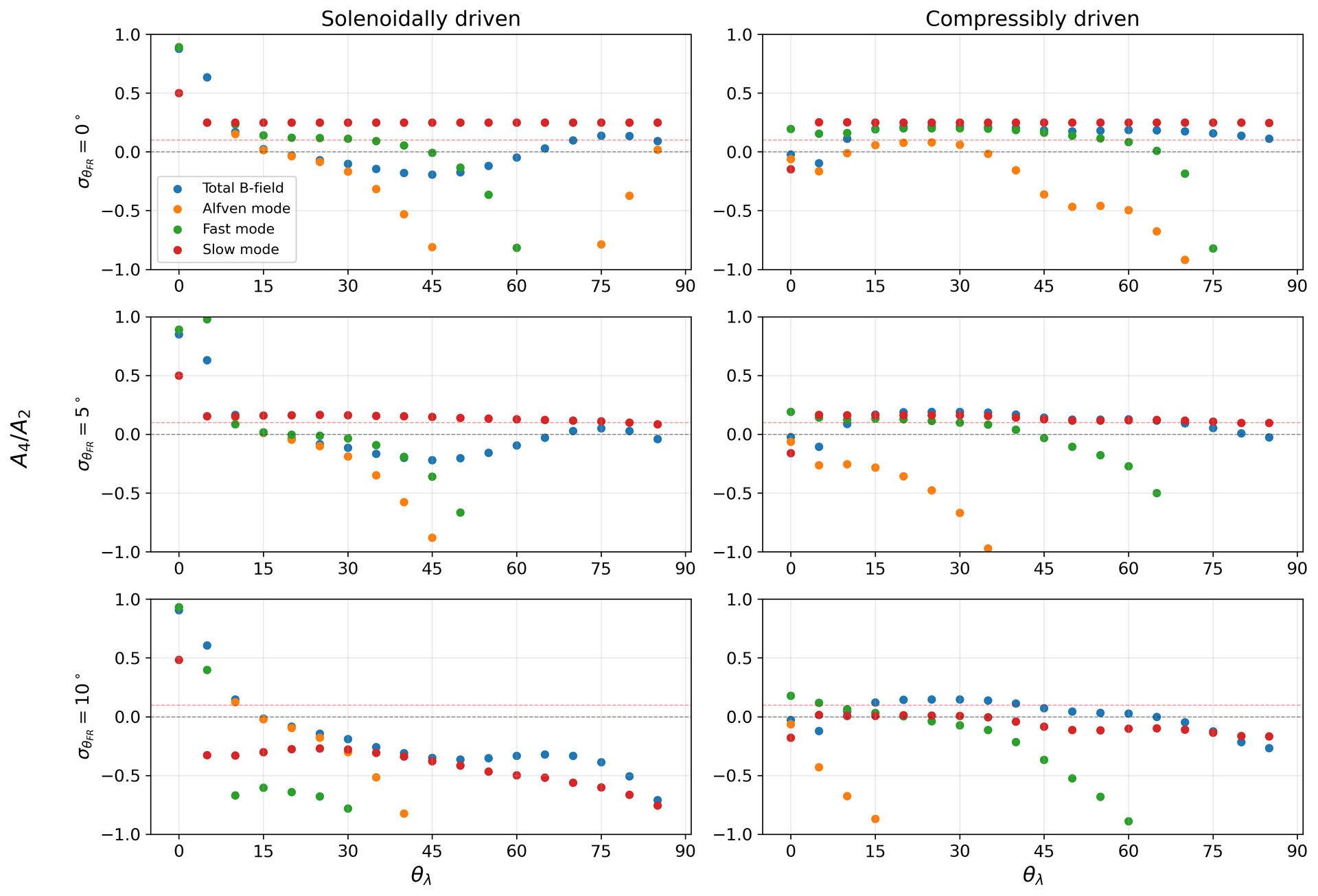}
    \caption{The parameter $A_4/A_2$ measured from synthetic synchrotron maps computed in the presence of a non-homogeneous foreground Faraday rotation screen. The top panels represent the case without FR while the middle and bottom panels show the cases for increasing inhomogeneity in the FR angles ($\sigma_{\theta_{FR}}=0^\circ,5^\circ, 10^\circ$ respectively). The left and right panels represent tests using synthetic maps from solenoidally (S7 )and compressively (C7) driven turbulence simulations respectively. The total magnetic field is shown in blue color, and the decomposed Alfv\'en, fast, and slow MHD modes are represented by the colors yellow, green, and red respectively. The red and black dashed lines show the upper and lower limits of the ambiguous classification region.}
    \label{fig:faraday_plot}
\end{figure*}

\end{document}